\documentclass[aps,prd,twocolumn,amsmath,amssymb,superscriptaddress,nofootinbib,floatfix]{revtex4-2}
\usepackage[T1]{fontenc}
\usepackage[utf8]{inputenc}
\IfFileExists{lmodern.sty}{\usepackage{lmodern}}{}
\usepackage{amsmath,amssymb,amsfonts}
\usepackage{graphicx}
\usepackage{booktabs}
\usepackage{xcolor}
\usepackage[colorlinks=true,linkcolor=blue!60!black,citecolor=blue!60!black,urlcolor=blue!60!black]{hyperref}

\newcommand{\trK}{\mathrm{tr}K}
\newcommand{\MADM}{M_{\rm ADM}}
\newcommand{\Mhor}{M_{\rm hor}}
\newcommand{\ThAH}{\Theta_{\rm AH}}
\newcommand{\wout}{w_{\rm out}}
\newcommand{\win}{w_{\rm in}}

\newcommand{\maybegraphic}[2][width=\textwidth]{%
  \IfFileExists{#2}{\includegraphics[#1]{#2}}{%
    \fbox{\parbox[c][4.5cm][c]{0.9\textwidth}{\centering\ttfamily
      [figure: \detokenize{#2}]}}}}

\begin{document}

\title{Gravitational collapse in Lorentz-violating gravity and the spin-0 mode}

\author{Emma Albertini}
\email{ealberti@sissa.it}
\affiliation{SISSA, Via Bonomea 265, 34136 Trieste, Italy and INFN Sezione di Trieste}
\affiliation{IFPU - Institute for Fundamental Physics of the Universe, Via Beirut 2, 34014 Trieste, Italy}

\author{Massimo Vaglio}
\affiliation{SISSA, Via Bonomea 265, 34136 Trieste, Italy and INFN Sezione di Trieste}
\affiliation{IFPU - Institute for Fundamental Physics of the Universe, Via Beirut 2, 34014 Trieste, Italy}
\author{Enrico Barausse}
\email{barausse@sissa.it}
\affiliation{SISSA, Via Bonomea 265, 34136 Trieste, Italy and INFN Sezione di Trieste}
\affiliation{IFPU - Institute for Fundamental Physics of the Universe, Via Beirut 2, 34014 Trieste, Italy}

\date{\today}

\begin{abstract}
Einstein--\ae ther theory is a theory of gravity with a preferred frame, in
which Lorentz symmetry is broken by a dynamical unit timelike vector field.
The theory is closely related to Ho\v{r}ava gravity, and in spherically
symmetric collapse the two produce the same dynamics. Both propagate a
scalar (spin-0) mode besides the tensor mode and, in Einstein--\ae ther
theory, the vector mode. The physical nature of the spin-0 mode has
recently been called into question following the post-Newtonian analysis of Taherasghari and Will, whose
post-Newtonian calculation of radiation reaction in compact binaries finds
no trace of its speed in the energy loss, suggesting that the mode may not
be a dynamical, energy-carrying wave. We test this with fully nonlinear
evolutions of scalar-field collapse in spherical symmetry, where the spin-0
mode is the only radiative degree of freedom. The collapse emits an outgoing
spin-0 wave that falls off as the inverse of the radius, travels at the
speed predicted by the linear theory, and carries positive energy to
infinity. The static \ae ther configuration left around the black hole has
negative energy instead, so that the horizon mass exceeds the ADM mass. For
large couplings, collapse does not produce a black hole: we recover the
transient white holes found by Akhoury, Garfinkle and Gupta, but at higher
resolution the collapsing shell keeps steepening, not hidden behind any
trapped surface, which may signal a naked singularity.
\end{abstract}
\maketitle

\section{Introduction}
\label{sec:intro}

Lorentz invariance has been tested to high precision in the matter
sector~\cite{Mattingly:2005re,Liberati:2013xla}. The bounds on its
violation in the gravitational sector are weaker, because there are few consistent frameworks in which such
violations can be parametrised beyond the weak
field~\cite{Jacobson:2008aj,Liberati:2013xla,Will:2014kxa}.
Einstein--\ae ther
theory~\cite{Gasperini:1987nq,Jacobson:2000xp,Jacobson:2008aj} provides
such a framework. It couples general relativity (GR) to a dynamical
vector field, the \ae ther, constrained to be timelike and of unit norm.
The \ae ther therefore selects a
preferred frame at every point, breaking local boost invariance, while preserving 
general covariance. The action contains every term
quadratic in the first derivatives of the \ae ther, weighted by four
dimensionless couplings, so the theory is the most general
two-derivative theory of this kind. The linear theory about flat space
propagates a spin-1 mode and a spin-0 mode besides the spin-2 mode of
GR, with speeds fixed by the couplings~\cite{Jacobson:2004ts}.\\
Lorentz-violating gravity is also motivated by approaches to quantum gravity.
Ho\v{r}ava~\cite{Horava:2009uw} formulated a theory on a preferred
foliation, with higher spatial derivatives but only second order in the time derivatives,
which scales anisotropically in the ultraviolet and is power-counting
renormalisable. Its projectable version, in which the lapse depends only on time, has been shown to be perturbatively
renormalisable~\cite{Barvinsky:2015kil}, whereas a general proof of perturbative renormalizability for the non-projectable theory is still lacking~\cite{Bellorin:2022qeu, Blas:2025ejp}. The consistent non-projectable
theory~\cite{Blas:2009qj,Blas:2010hb} contains a scalar field with
timelike gradient, the khronon, whose level sets are the leaves of the
foliation. Its low-energy limit is Einstein--\ae ther theory with the \ae
ther constrained to be the unit normal to the khronon
foliation~\cite{Jacobson:2010mx,Jacobson:2013xta}. In spherical symmetry
the \ae ther is necessarily hypersurface orthogonal, so every spherically
symmetric solution of Einstein--\ae ther theory also solves the
low-energy Ho\v{r}ava equations. The converse also holds, when
regularity at the universal horizon or at the centre is imposed~\cite{Jacobson:2010mx,Blas:2010hb,Barausse:2011pu,Blas:2011ni,Bhattacharyya:2015uxt}.\\
The couplings of both theories are constrained by theory and by
observation. Stability of the linear modes requires positive squared
speeds~\cite{Jacobson:2004ts} and positive
energy~\cite{Eling:2005zq,Garfinkle:2011iw}. The observation of
ultra-high-energy cosmic rays requires that no gravitational mode
propagate slower than light, since the cosmic rays would otherwise lose
their energy to gravitational Cherenkov radiation~\cite{Elliott:2005va}.
In the solar system the theory differs from GR only through the
preferred-frame parameters $\alpha_1$ and $\alpha_2$ of the parametrised
post-Newtonian formalism~\cite{Foster:2005dk}, which are tightly
bounded~\cite{Will:2014kxa}. The \ae ther rescales the gravitational
constant that drives the cosmological expansion relative to the one
measured locally. The primordial abundances of light elements bound this
rescaling~\cite{Carroll:2004ai}, and the cosmic microwave background and
large-scale structure give further bounds~\cite{Audren:2014hza}. Compact
objects acquire a dependence of their mass on their velocity relative to
the \ae ther, measured by their
sensitivities~\cite{Foster:2007gr,Yagi:2013qpa,Yagi:2013ava}. This
violates the strong equivalence principle and produces dipole
gravitational
radiation~\cite{Foster:2006az,Foster:2007gr,Yagi:2013qpa,Yagi:2013ava}.
The orbital decay of binary pulsars is sensitive to this dipole, and
their timing has become the main probe of the
couplings~\cite{Foster:2007gr,Yagi:2013qpa,Yagi:2013ava,Gupta:2021vdj,Vaglio:2026vmo}.
Finally, the coincident detection of GW170817 and
GRB~170817A~\cite{Abbott:2017gw} fixed the speed of the spin-2 mode to
that of light to within a part in $10^{15}$. This requires the combination of the coupling parameters $c_1+c_3$ to
vanish~\cite{Oost:2018tcv,EmirGumrukcuoglu:2017cfa,Ramos:2018oku,Barausse:2019yuk}. The surviving parameter space is significantly restricted, but some directions remain weakly constrained due to parameter degeneracies~\cite{Oost:2018tcv,Gupta:2021vdj,Barausse:2019yuk}.

These theories also alter the causal structure of black holes. Each mode propagates along the null cone of its own effective metric and is therefore associated with a distinct horizon. The horizon of a faster mode lies inside that of a slower one, so that a
stationary black hole is bounded by a nested set of horizons. Static
spherical solutions are singled out by regularity at the spin-0 horizon~\cite{Eling:2006ec,Barausse:2011pu},
slowly rotating ones have been obtained to first order in the
spin~\cite{Barausse:2015frm}, and rapidly rotating ones have been
constructed numerically by imposing regularity at each mode
horizon~\cite{Adam:2021vsk}. In Ho\v{r}ava
gravity the higher spatial derivatives make the dispersion relations
anisotropic, $\omega^2=k^2+\alpha k^4+\beta k^6$ schematically, so the
group velocity grows without bound in the ultraviolet and no finite speed
limits the propagation of signals~\cite{Blas:2011ni,Bhattacharyya:2015uxt}.
Static solutions nevertheless contain a universal horizon~\cite{Barausse:2011pu,Blas:2011ni}: a leaf of
the preferred foliation at constant radius, located inside the mode
horizons, which traps even arbitrarily fast signals~\cite{Berglund:2012bu,Bhattacharyya:2015gwa,Barausse:2013nwa}.
The dynamical properties of these surfaces and whether they form during gravitational collapse cannot be determined from stationary solutions alone.
 \\
 
Gravitational collapse in these theories has been studied mostly in
spherical symmetry. Garfinkle, Eling and Jacobson
(GEJ)~\cite{Garfinkle:2007bk} evolved a massless scalar field, initially
a spherical shell, in Einstein--\ae ther theory.
For moderate couplings, a black hole with a regular spin-0 horizon formed
and settled to the static solution of Ref.~\cite{Eling:2006ec}; at large
couplings, the fields developed steep gradients, which GEJ read as a possible
naked singularity. Akhoury, Garfinkle and Gupta~\cite{Akhoury:2016mrw}
revisited this case using both the GEJ gauge and a second, zero-shift radial gauge,
ingoing initial data and improved boundary conditions, and found instead a transient white hole: an anti-trapped
surface that forms and later disperses. Bhattacharyya, Coates, Colombo and
Sotiriou~\cite{Bhattacharyya:2015uxt} showed that in spherical symmetry,
with regularity at the centre or at the universal horizon, hypersurface-orthogonal Einstein--\ae ther
theory and low-energy Ho\v{r}ava gravity have identical dynamics, so the
GEJ simulations also describe Ho\v{r}ava collapse. They interpreted the
late-time vanishing of the preferred-frame lapse in those simulations as
the dynamical formation of a universal horizon. Such formation had been
shown for a thin shell collapsing in the cuscuton
model~\cite{Saravani:2013kva}, a scalar field with infinite sound speed
that coincides with a special case of low-energy Ho\v{r}ava gravity, and
was later found in the version of Ho\v{r}ava gravity that survives the
GW170817 and solar-system bounds~\cite{Franchini:2021bpt}, where the
metric collapses exactly as in GR while the khronon foliation asymptotes
to a limiting leaf. Whether universal horizons form generically, and whether gravitational collapse excites additional propagating gravitational degrees of freedom, are therefore central questions in these theories, and motivate the present work.

A related question concerns the dynamical status of the spin-0
mode. Recently Taherasghari and Will~\cite{Taherasghari:2023rwn,Taherasghari:2025mlf}
applied the post-Minkowskian expansion and the direct integration of the relaxed field equations to Einstein--\ae ther theory, two methods that
underlie high-order post-Newtonian results in GR. Since the unit-norm
constraint spoils the usual form of the relaxed equations, they redefined
the metric and \ae ther variables to obtain flat-space wave equations with
sources quadratic in the fields, solved them in the near zone to 2.5
post-Newtonian order~\cite{Taherasghari:2023rwn}, and derived the
equations of motion of a compact binary, including radiation reaction at
1.5 and 2.5 post-Newtonian order, and its energy
loss~\cite{Taherasghari:2025mlf}. They found that the dipole term, the
leading effect in the orbital decay, depends only on the speed $v_T$ of
the transverse \ae ther waves. Previous works by Foster~\cite{Foster:2006az,Foster:2007gr}
and Yagi, Blas, Barausse and Yunes~\cite{Yagi:2013ava} instead computed
the far-zone energy flux, and found that the dipole flux
depends also on the speed $v_L$ of the longitudinal mode, i.e.\ the
spin-0 speed $v_0$ of Eq.~\eqref{eq:speeds} below. In the relaxed equations of Ref.~\cite{Taherasghari:2025mlf}, by contrast, the metric potentials propagate at the speed of light, the \ae ther potential at $v_T$, and no wave equation carries $v_L$. 

The authors emphasize that the two calculations are conceptually distinct: one describes radiation reaction in the near zone, while the other yields the radiative flux in the far zone. Their equivalence relies on the energy-balance argument, which has been subject to only limited explicit verification even in GR and whose validity in alternative theories of gravity cannot be taken for granted. Consequently, they leave open whether energy balance fails in this theory or one of the two approaches is flawed. Their result, however, raises a further question: is the spin-0 mode a dynamical, energy-carrying wave propagating at its own speed, as the linear analysis indicates~\cite{Jacobson:2004ts,Eling:2005zq}, or is its apparent contribution to the radiated energy an artefact of the far-zone calculation? In the latter case, the dipole bounds~\cite{Foster:2007gr,Yagi:2013qpa,Yagi:2013ava,Gupta:2021vdj,Vaglio:2026vmo}
would have to be revisited, and the theory would differ qualitatively
from its linearized description~\cite{Jacobson:2004ts,Foster:2006az}.

Fully nonlinear spherical collapse is a natural place to address these
questions. In spherical symmetry the spin-2 and spin-1 modes are absent,
so the spin-0 mode is the only gravitational degree of freedom that can
radiate, and its fall-off, speed and energy can be measured directly on
the full solution rather than in a post-Newtonian expansion. We therefore
return to the GEJ problem, starting from their couplings and extending
them to a superluminal spin-0 speed, which separates the spin-0 wave from
the scalar field, and to the large-$c_1$ case of Akhoury, Garfinkle and
Gupta. These couplings lie far outside the observational bounds and serve
as a laboratory for the dynamics of the theory rather than its
phenomenology. Our results show that the spin-0 mode propagates dynamically at its characteristic speed and carries positive energy to infinity. This provides a different picture from that suggested by the post-Newtonian analysis of Refs.~\cite{Taherasghari:2023rwn,Taherasghari:2025mlf}, where the longitudinal mode does not appear in the radiation-reaction terms. We return to this apparent tension and discuss a possible origin of the discrepancy in Sec.~\ref{sec:pn_comparison}.

The paper is
organised as follows. Section~\ref{sec:framework} reviews Einstein--\ae
ther theory, the GEJ formulation, our numerical method, and the
couplings, initial data and diagnostics used. Section~\ref{sec:conv}
establishes the convergence of the free evolution.
Section~\ref{sec:mass} studies the Misner--Sharp mass profile and the
horizon. Section~\ref{sec:spin0} identifies the spin-0 mode and measures
its fall-off, its speed and the energy it carries and discusses the tension with the post-Newtonian results. Section~\ref{sec:wh}
treats collapse at large couplings and Section~\ref{sec:conclusion} summarizes
the results. Appendix~\ref{sec:app-conv} collects the convergence
tests while Appendix~\ref{app:pn_redefinition} provides the technical details underlying the argument presented in Sec.~\ref{sec:pn_comparison}.
We use units with $G=c=1$ unless otherwise stated, and adopt a metric signature $(-,+,+,+)$.
The code, the scripts that produce all figures and numbers of this paper,
and symbolic checks of the equations are publicly
available~\cite{AEcollapse_code}.
\\

\section{Einstein--\ae ther theory}
\label{sec:framework}

Einstein--\ae ther theory is general relativity (GR) coupled
to a dynamical, unit-norm, timelike vector field $u^a$, the \ae
ther~\cite{Jacobson:2000xp,Jacobson:2008aj,Berti:2015itd,Sarbach:2019yso}. Because $u^a$ is nowhere
vanishing, it selects a preferred frame at every point and breaks local
Lorentz invariance while keeping the theory generally covariant. It is
the most general such theory with at most two derivatives acting on the
fields. The action is
\begin{equation}
  \begin{split}
  S={}&\frac{1}{16\pi G}\int d^4x\,\sqrt{-g}\,
  \Bigl(R - M^{ab}{}_{mn}\nabla_a u^m\nabla_b u^n\\
  &+ \lambda\,(u^a u_a+1)\Bigr) + S_{\rm matter},
  \end{split}
  \label{eq:action}
\end{equation}
with
\begin{equation}
  M^{ab}{}_{mn}=c_1 g^{ab}g_{mn}+c_2\delta^a_m\delta^b_n
  +c_3\delta^a_n\delta^b_m-c_4u^au^bg_{mn},
\end{equation}
 and a Lagrange multiplier $\lambda$ that enforces
the unit-norm condition. The four dimensionless couplings $c_i$ are also indicated through the combinations $c_{ij}\equiv c_i+c_j$ and
$c_{123}\equiv c_1+c_2+c_3$. When the \ae ther is hypersurface-orthogonal,
as it necessarily is in spherical symmetry, the twist term can be
removed and the theory coincides with the infrared limit of Ho\v{r}ava
gravity~\cite{Horava:2009uw,Blas:2009qj,Jacobson:2010mx,Blas:2010hb,Jacobson:2013xta,Franchini:2021bpt,Barausse:2019yuk}.

Linearised about flat space with a constant \ae ther, the theory has five
propagating degrees of freedom: two spin-2, two spin-1 and one spin-0
mode~\cite{Jacobson:2004ts}, with squared speeds
\begin{equation}
  \begin{gathered}
  v_2^2=\frac{1}{1-c_{13}},\qquad
  v_1^2=\frac{2c_1-c_1^2+c_3^2}{2c_{14}(1-c_{13})},\\
  v_0^2=\frac{c_{123}(2-c_{14})}{c_{14}(1-c_{13})(2+c_{13}+3c_2)} .
  \end{gathered}
  \label{eq:speeds}
\end{equation}
The energy densities of the linear modes were computed by
Eling~\cite{Eling:2005zq} with the Weinberg and Einstein energy--momentum
pseudotensors; that of the spin-0 mode has the sign of
$c_{14}(2-c_{14})$~\cite{Eling:2005zq,Jacobson:2008aj}, so the spin-0
mode carries positive energy only if $0<c_{14}<2$. In spherical
symmetry only the spin-0 mode survives%
: it is the only channel
through which the spacetime can radiate apart from the matter itself.

Observationally, the parameter space is now tightly constrained~\cite{Foster:2005dk,Elliott:2005va,Foster:2007gr,Yagi:2013ava,Ramos:2018oku,Vaglio:2026vmo}.
The coincident detection of GW170817 and GRB~170817A~\cite{Abbott:2017gw}
bounds the spin-2 speed and hence $|c_{13}|\lesssim10^{-15}$~\cite{Oost:2018tcv,EmirGumrukcuoglu:2017cfa,Gupta:2021vdj}.
The couplings used in this paper are outside this range. We
choose them deliberately, as a numerical laboratory: two of them are the
coupling sets of Refs.~\cite{Garfinkle:2007bk,Akhoury:2016mrw}, allowing a
direct comparison, the third gives a superluminal spin-0 mode that
separates from the scalar field, and all of them make the \ae ther sector strongly
nonlinear, which is the hardest test of the scheme. Our results should
be read as statements about the dynamics of the theory, not as
predictions for viable couplings.

\section{Collapse in Einstein--\ae ther theory}

Black holes in Einstein--\ae ther theory have a richer causal structure than in GR:
each mode has its own horizon, and the stationary solutions are fixed by
regularity at the spin-0 horizon~\cite{Eling:2006ec,Eling:2006df,Barausse:2011pu,Blas:2011ni,Berglund:2012bu,Adam:2021vsk,Barausse:2013nwa,Barausse:2015frm,Saravani:2013kva}.
We build on GEJ~\cite{Garfinkle:2007bk}, who showed that
such black holes form dynamically in scalar-field collapse. Using the GEJ formulation and a second radial gauge,
Akhoury, Garfinkle and Gupta~\cite{Akhoury:2016mrw} later argued that at
large $c_1$, collapse can produce a transient white hole rather than a
black hole. In the following, we review our numerical formulation and test the convergence of our set up.

\subsection{Formulation and numerical method}
\label{sec:method}

Since in spherical symmetry the \ae ther is hypersurface orthogonal, 
 GEJ~\cite{Garfinkle:2007bk} write it as
 $u_a=-N\nabla_a t$, i.e.,
the spacetime foliation is chosen to match the \ae ther-orthogonal slices.
This is known as unitary gauge~\cite{Blas:2010hb}. The lapse 
is therefore not a gauge variable, but is fixed by the \ae ther through the
kinematic identity
\begin{equation}
  a_i=D_i\ln N ,\qquad a^a\equiv u^b\nabla_b u^a .
  \label{eq:lapse}
\end{equation}
We work in spherical symmetry, with a massless scalar field $\psi$
normalised as in GEJ~\cite{Garfinkle:2007bk}, so that the field
equations read $G_{ab}=T_{ab}$, i.e.\ $\psi$ is $\sqrt{8\pi}$ times a
canonically normalised field. The spacetime metric can be written as
\begin{equation}
  ds^2=-N^2(t,r)\,dt^2+(dr+\beta(t,r)\,dt)^2
  +\Theta^2(t,r)\,d\Omega^2 ,
  \label{eq:metric}
\end{equation}
where $\beta(t,r)$ is the radial shift and $\Theta(t,r)$ is the areal
radius, defined such that the symmetry spheres have area
$4\pi\Theta^2$. We denote by $\gamma_{ij}$ the spatial metric induced
on the \ae ther-orthogonal slices, which determines the proper spatial
line element $d\ell^2=\gamma_{ij}dx^i dx^j$. In this form of the metric
$r$ measures proper radial distance along the slices, i.e.\
$\gamma_{rr}=1$.

Following GEJ~\cite{Garfinkle:2007bk}, the freely evolved variables are
\begin{equation}
  \bigl(\psi,\,P,\,\trK,\,\Theta,\,a_r\bigr),
\end{equation}
namely the scalar, its momentum $P\equiv u^a\nabla_a\psi$ along the
\ae ther, the trace of the extrinsic curvature,
the areal radius, and the radial \ae ther acceleration. The shear $Q$ of
the extrinsic curvature, defined by $K^r{}_r=\trK/3+Q$ and
$K^\theta{}_\theta=K^\phi{}_\phi=\trK/3-Q/2$, the lapse $N$ and the
shift $\beta^r\equiv\beta$ are recovered on each slice. The 3+1 field
equations in unitary gauge, for general matter and without symmetry, are
given in Ref.~\cite{Bonetti:2015oda} (their $\alpha$, $\beta$, $\lambda$
are our $c_{14}$, $c_{13}$, $c_2$, with $f(a)=\alpha a^2$). In spherical
symmetry, with a prime for $\partial_r$, the evolution equations are
\begin{align}
  (\partial_t-\beta\partial_r)\psi&=NP,\nonumber\\
  (\partial_t-\beta\partial_r)\Theta&=N\Theta\Bigl(\frac Q2-\frac{\trK}{3}\Bigr),\nonumber\\
  (\partial_t-\beta\partial_r)P&=N\Bigl[\trK P+\psi''+\Bigl(a_r+\frac{2\Theta'}{\Theta}\Bigr)\psi'\Bigr],\nonumber\\
  (\partial_t-\beta\partial_r)\trK&=\frac N3\,\trK^2\nonumber\\
  &\quad+\frac{N}{2+c_{13}+3c_2}\Bigl[2P^2+3(1-c_{13})Q^2\nonumber\\
  &\quad+(c_{14}-2)\Bigl(a_r'+a_r^2+\frac{2\Theta'}{\Theta}a_r\Bigr)\Bigr],\label{eq:evolution}\\
  (\partial_t-\beta\partial_r)a_r&=N\Bigl(\frac{2\trK}{3}-Q\Bigr)a_r\nonumber\\
  &\quad+N\,\frac{c_{13}P\psi'-c_{123}\trK'}{c_{14}(1-c_{13})}.\nonumber
\end{align}
The variables $Q$, $N$ and $\beta$ follow from
\begin{equation}
  \begin{gathered}
  (1-c_{13})\Bigl(Q'+\frac{3\Theta'}{\Theta}Q\Bigr)
   =\frac{2+c_{13}+3c_2}{3}\,\trK'-P\psi',\\
  \beta'=N\Bigl(Q+\frac{\trK}{3}\Bigr),
  \end{gathered}
  \label{eq:radial}
\end{equation}
and Eq.~\eqref{eq:lapse}, integrated outward from $Q(0)=0$ and from the
regularity condition $\beta(0)=0$, and inward from $N\to1$ at large
radius. The Hamiltonian constraint is not imposed; multiplied by
$\Theta$, it reads
\begin{equation}
  \begin{split}
  \Theta C\equiv{}&\Theta\Theta''+\tfrac12(\Theta'^2-1)+c_{14}\,a_r\Theta\Theta'\\
  &+\tfrac{\Theta^2}{4}\Bigl[c_{14}(2a_r'+a_r^2)+P^2+\psi'^2\\
  &+\tfrac32(1-c_{13})Q^2-\tfrac13(2+c_{13}+3c_2)\trK^2\Bigr]=0 ,
  \end{split}
  \label{eq:hamiltonian}
\end{equation}
and we monitor it during the evolution.

\emph{Free evolution of $a_r$.} Although $a_r$ is fixed on each slice
by the lapse through Eq.~\eqref{eq:lapse}, we evolve it with the radial
\ae ther equation combined with the momentum
constraint~\cite{Garfinkle:2007bk}. Because in unitary gauge the
\ae ther equation follows from the metric
equations~\cite{Jacobson:2011cc}, this keeps the Hamiltonian constraint
satisfied as long as $N$ is rebuilt from $a_r$ with
Eq.~\eqref{eq:lapse}. We check this by relaxing Eq.~\eqref{eq:lapse} to
$\partial_r\ln N=\bar\alpha\,a_r$. Near the centre, the constraint of
Eq.~\eqref{eq:hamiltonian} behaves as $C=C_1\,r+O(r^3)$, and for data
with $C=0$
\begin{equation}
  \left.\frac{dC_1}{dt}\right|_{C=0}
  =(1-\bar\alpha)\,a_r'(0)\,\trK(0)\,N(0),
  \label{eq:constraint-prop}
\end{equation}
which vanishes only for $\bar\alpha=1$.

\emph{The slicing is singularity avoiding.} Integrating
Eq.~\eqref{eq:lapse} inward from $N=1$ at large radius gives
$\ln N(r)=-\int_r^\infty a_r\,dr'$. Once the horizon has formed, $a_r$ is
non-negative inside it, so $N$ decreases monotonically inward there. The
lapse does not collapse at the apparent horizon, where it stays of order
unity, but at a smaller radius inside it, where it drops by many orders
of magnitude over a narrow range of $\Theta$ (by $t\simeq150$ this happens
at $\Theta\simeq8$, with $\ThAH\simeq12$); inside that radius it is
exponentially small all the way to the centre. Every source term in the
evolution equations is proportional to $N$. The advection terms are
proportional to $\beta$ instead, but Eq.~\eqref{eq:radial}
and $\beta(0)=0$ together with the monotonicity of $N$ give
\begin{equation}
  |\beta(r)|\le N(r)\int_0^r\Bigl|Q+\frac{\trK}{3}\Bigr|\,dr' ,
\end{equation}
so $\beta$ also vanishes wherever the lapse has collapsed. The fields
inside that radius therefore freeze before the singularity is reached,
as in the original GEJ simulations~\cite{Garfinkle:2007bk}. 

This freezing has a geometric
interpretation. In static black holes the leaves of the \ae ther
foliation do not cross the universal horizon but accumulate on it from
outside, so the preferred time diverges there and the lapse of the \ae
ther slicing vanishes~\cite{Barausse:2011pu,Barausse:2013nwa}.
Accordingly, Ref.~\cite{Bhattacharyya:2015uxt} reads the vanishing of the
lapse in the GEJ simulations as the dynamical formation of a universal
horizon, which in Ho\v{r}ava gravity is also a Cauchy
horizon~\cite{Bhattacharyya:2015gwa}. On this reading the frozen region is
the interior of the forming universal horizon, which the \ae ther slicing
cannot penetrate. The freezing must not spread to the
region we want to study, and we check this directly in the baseline
collapse run.
 The lapse at the apparent horizon
stays of order unity, so the horizon
and everything outside it continue to evolve and the growth of the
horizon can be followed to late times.
When the interior freezes while the
exterior keeps moving, slices can stretch, building steep gradients at
the edge of the frozen region that eventually cannot be resolved. Here
$\max|a_r|$ saturates instead of growing, so the gradients stay bounded.
Because the singularity is never reached, there is nothing to excise:
the grid extends to $r=0$ throughout, with no interior boundary.

\begin{figure*}[t]
\centering
\maybegraphic[width=\textwidth]{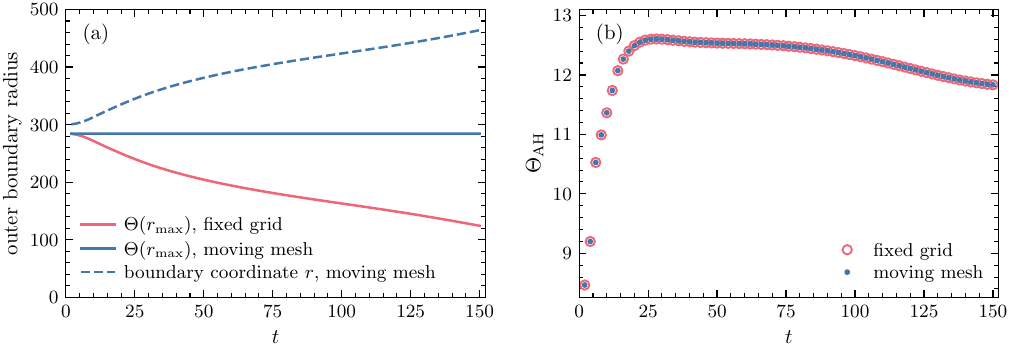}
\caption{Moving-mesh validation ($r_{\max}=300$, $N_r=25600$, $A_0=0.35$).
(a) Outer boundary against $t$. On the fixed grid, the areal radius of
the outer boundary, $\Theta(r_{\max})$, shrinks with time
(red); the moving mesh holds it
fixed (blue, solid) by advecting the boundary coordinate $r$ outward (blue,
dashed). (b) The areal radius $\ThAH$ of the apparent horizon against
$t$ on the fixed grid (red open circles) and on the moving mesh (blue
dots); the two
agree, showing that the map only changes
the coordinates. Note that $\ThAH$ reaches its maximum at
$t\simeq28$ and then slowly decreases, as discussed in
Sec.~\ref{sec:mass}.}
\label{fig:mesh}
\end{figure*}

\emph{Discretization.} Spatial derivatives are second-order centered finite
differences on a uniform grid of $N_r$ points,
$dr=r_{\max}/(N_r-1)$, with regularity at $r=0$ imposed through
parity; time integration is by iterated Crank--Nicolson~\cite{Teukolsky:1999rm}, with fourth-order
Kreiss--Oliger dissipation~\cite{KreissOliger}. At the outer boundary,
conditions are imposed only on the incoming characteristic fields of the
spin-0 mode and of the scalar, and only while these fields enter the grid.
This treatment relies on the characteristic analysis of the spin-0
sector and is described in Sec.~\ref{sec:infinity}. Because, in the $\gamma_{rr}=1$ gauge, the grid falls inward
($\beta$ is nearly uniform across the exterior and of order $-1$), the outer boundary shrinks in areal
radius at roughly one areal unit per unit time
(Fig.~\ref{fig:mesh}(a)). For the long runs of
Sec.~\ref{sec:spin0} we therefore use an arbitrary Lagrangian--Eulerian~\cite{Hirt:1974}
map $r=R(x,t)=x+g(x)s(t)$, with $g$ a $C^2$ quintic taper supported for
$x>0.4\,r_{\max}$ (the outer $60\%$ of the grid) and $s(t)$ chosen to
hold $\Theta(r_{\max})$ fixed.
This changes coordinates only: with the map switched off, the code
reproduces the fixed-grid run, and with it switched on, the
apparent-horizon radius agrees with the fixed grid (Fig.~\ref{fig:mesh}(b)).

\emph{Couplings and initial data.} Unless stated otherwise,
$c_1=0.7$, $c_2=-0.5119403$, $c_3=c_4=0$ (so $c_{13}=c_{14}=0.7$), the
values of GEJ~\cite{Garfinkle:2007bk} for which $v_0=1$. These will be referred to as baseline couplings.  To separate the spin-0 signal from the scalar, we also study the case
$c_2=0$ ($v_0=1.267$). The initial data are a time-symmetric scalar
shell, a variant of that of GEJ~\cite{Garfinkle:2007bk} (whose profile is
$a_0\exp[-(r^2-r_0^2)^2/s^4]$) multiplied by $r^2/100$,
\begin{equation}
  \psi(0,r)=\frac{A_0}{100}\,r^2\exp\!\Bigl[-\Bigl(\frac{r^2-r_0^2}{\sigma^2}\Bigr)^{2}\Bigr],
  \qquad P(0,r)=0,
  \label{eq:id}
\end{equation}
with $\trK=a_r=0$ as in GEJ~\cite{Garfinkle:2007bk} (hence $Q=0$), and
$r_0=10$ and $\sigma=4$ (the GEJ values), a shell centred at $r\simeq10$
with peak value $\simeq A_0$. At the
baseline amplitude $A_0=0.35$ it collapses to a black hole.

\emph{Diagnostics and physical quantities.} We use the expansions of
the outgoing and ingoing radial null congruences orthogonal to the
symmetry spheres, with null normals $u^a\pm e^a$ ($e^a$ the unit radial
vector), rescaled by $\Theta/2$ and with the sign of the ingoing one
reversed, so that both are positive on an ordinary sphere:
\begin{equation}
  \theta_\pm\equiv\partial_r\Theta\pm\tfrac16(3Q-2\trK)\Theta .
  \label{eq:expansions}
\end{equation} The
Misner--Sharp mass~\cite{MisnerSharp,Hayward:1994bu} is
\begin{equation}
  m=\frac{\Theta}{2}\,\bigl(1-\theta_+\theta_-\bigr),
  \label{eq:ms}
\end{equation}
the apparent horizon is the outermost surface where $\theta_+=0$, and the
ADM mass is read where $m(\Theta)$ is flat. Because the \ae ther energy
density has a long negative tail, this requires $\Theta\gtrsim250$, beyond the support of the scalar.
On the apparent horizon $\theta_+=0$, so Eq.~\eqref{eq:ms} gives
$m=\ThAH/2$: the horizon mass $\Mhor\equiv m(\ThAH)=\ThAH/2$ is fixed by
the areal radius of the horizon, and the horizon mass and the horizon area
$4\pi\ThAH^2$ grow or shrink together.

\subsection{Convergence}
\label{sec:conv}

We test the convergence of the code in two regimes. In the first,
scattering, a shell falls through the regular centre, bounces and
disperses without forming a horizon. This tests the formulation and the
treatment of the centre in a spacetime that is nonlinear but regular
everywhere. In the second, collapse, the shell forms a black hole and the
interior freezes. This tests the region outside the horizon, from which
the results of the following sections are extracted.

Both tests follow the same procedure. We evolve the same initial data on
three grids, with spacings $h$, $h/2$ and $h/4$, and compare the results.
The grids are nested: every point of the coarsest grid is also a point of
the two finer ones, so that the differences between resolutions are
taken pointwise, with no interpolation. For this reason both tests use
the fixed grid, without the moving mesh. The convergence order for a quantity $f$ is
\begin{equation}
  p=\log_2\frac{\|f_{h}-f_{h/2}\|_2}{\|f_{h/2}-f_{h/4}\|_2},
\end{equation}
where $f_h$ is $f$ computed with grid spacing $dr=h$ and time step
$dt=0.1\,h$, so that both are halved together, and $\|\cdot\|_2$ is the
$L_2$ norm over $r$ on the coarsest grid, in a window given below for
each test. The scheme is second order in space and time, so the error
of $f_h$ should scale as $h^2$, and the expected result is $p=2$.
Equivalently, the difference between the two finer grids, multiplied by
$4$, should coincide with the difference between the two coarser ones.
The figures of Appendix~\ref{sec:app-conv} show the norms in this form.

The Hamiltonian constraint provides an independent check. It vanishes
for the exact solution, so its value on the grid is itself a measure of
the numerical error, and its norm should decrease by a factor $4$ each
time $dr$ is halved. Its order is therefore $\log_2$ of the ratio of its
norms on two successive grids, averaged over the two pairs of grids.
The constraint $C$ itself contains the term $(\Theta'^2-1)/(2\Theta)$,
which is $0/0$ at a regular centre and cannot be evaluated reliably there
on the grid. We therefore monitor $\Theta C$, Eq.~\eqref{eq:hamiltonian},
which is finite at $r=0$: on a regular slice $\Theta$ and $a_r$ are odd
in $r$ and the other fields even, so near the centre every term of
$\Theta C$ vanishes except $\tfrac12(\Theta'^2-1)$, and
$\Theta C\to\tfrac12[\Theta'(0)^2-1]$. The constraint at $r=0$ is then
the regularity condition $\Theta'(0)=1$.

\emph{Scattering.} We evolve the initial data of Eq.~\eqref{eq:id} with
three amplitudes, $A_0=0.01$, $0.04$ and $0.06$, on the grids
$N_r=6401/12801/25601$ ($r_{\max}=150$, $dr=0.023/0.012/0.006$). In all
three cases the shell falls through the centre, bounces and disperses.
The field grows stronger with $A_0$. We measure its strength by the
minimum of the central lapse during the bounce, which is
$N_{\min}\simeq0.98$, $0.70$ and $0.4$ for the three amplitudes. We do
not go beyond $A_0=0.06$. For stronger shells the coarsest grid no longer
resolves the bounce, and a trapped region a few cells wide appears at the
centre, a resolution artefact discussed in Sec.~\ref{sec:wh}. We compute
the norms over $0<r<20$, which contains the centre and the outgoing
shell. We average the orders over $t=12$--$14$, just after the bounce.
Table~\ref{tab:conv-scatter} lists the orders, and
Appendix~\ref{sec:app-conv} shows the norms against time.

\begin{table*}[t]
\centering
\caption{Measured convergence order $p$ (target $2$). First three rows:
scattering, nested grids $N_r=6401/12801/25601$ ($r_{\max}=150$), $L_2$
norm over $0<r<20$, orders averaged over $t=12$--$14$; $N_{\min}$ is the
minimum of the central lapse. Last row: collapse, nested grids
$N_r=12801/25601/51201$ ($r_{\max}=300$), $L_2$ norm outside the
apparent horizon, over $r_{\rm AH}+2\le r\le150$, orders averaged over
$t=4$--$28$.}
\label{tab:conv-scatter}
\begin{tabular}{lccccccc}
\toprule
 & $\psi$ & $P$ & $\trK$ & $a_r$ & $\Theta$ & $Q$ & $\|\Theta C\|$\\
\midrule
$A_0=0.01$ (weak)                       & 2.35 & 2.52 & 2.17 & 2.16 & 2.92 & 2.16 & 1.95\\
$A_0=0.04$ (strong, $N_{\min}\approx0.70$) & 2.62 & 2.46 & 2.20 & 2.14 & 3.36 & 2.12 & 1.90\\
$A_0=0.06$ (strongest, $N_{\min}\approx0.4$) & 2.81 & 1.99 & \textbf{1.45} & \textbf{1.47} & 2.42 & \textbf{1.42} & \textbf{1.49}\\
\midrule
Collapse, $A_0=0.35$                    & 2.62 & 2.55 & 2.34 & 2.38 & 2.62 & 2.42 & 2.24\\
\bottomrule
\end{tabular}
\end{table*}

For $A_0=0.01$ and $0.04$, all fields and the constraint converge at
second order. For $A_0=0.06$, the scalar field and the areal radius
still converge at second order. The \ae ther variables $\trK$, $a_r$ and
$Q$ and the constraint, instead, converge only with $p\approx1.4$--$1.5$.
This loss of order appears after the bounce, when the shell is most
compressed at the centre, and our coarsest grid no longer fully resolves
it. The convergence indeed improves with resolution. At $t=14$, the
constraint norm decreases by a factor $1.8$ between the two coarser grids
and by $2.5$ between the two finer ones, approaching the factor $4$ of
second order.

\begin{figure}[htbp]
\centering
\maybegraphic[width=\columnwidth]{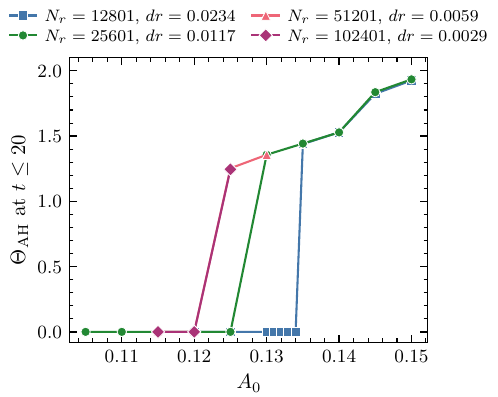}
\caption{Black-hole formation threshold for the time-symmetric shell
($c_1=0.7$, $r_0=10$, $\sigma=4$) on four grids: areal radius of the
apparent horizon that forms when the shell reaches the centre ($t\le20$)
against the amplitude of the initial pulse, set to zero where no horizon
forms.}
\label{fig:threshold}
\end{figure}

All three scattering amplitudes lie well below the black-hole threshold
$A_*$. We locate it by bisection in $A_0$ on four grids, from
$N_r=12801$ ($dr=0.023$) to $N_r=102401$ ($dr=0.003$). We identify
black-hole formation with the appearance of an apparent horizon of
finite areal radius when the shell reaches the centre
(Fig.~\ref{fig:threshold}). Unlike the trapped region a few cells wide
that the coarsest grid shows for stronger scattering shells, this horizon
has the same areal radius on every grid that forms it. The smallest
amplitude that forms it decreases with resolution, from $0.135$ on the
coarsest grid to $0.125$ on the two finest. Therefore $A_*$ lies between
$0.120$ and $0.125$. Above $A_*$, the areal radius of the horizon grows
with $A_0$.

\emph{Collapse.} For the collapse test we take the baseline shell,
$A_0=0.35$, on the grids $N_r=12801/25601/51201$ with $r_{\max}=300$, so
that $dr=0.023/0.012/0.006$ as in the scattering test, and compare the
three runs every two time units up to $t=28$. An apparent horizon is
already present at $t=2$. Inside it, the lapse collapses and the fields
freeze, as described in Sec.~\ref{sec:method}. The frozen interior
develops steep gradients that a fixed grid cannot resolve. It does not
affect the exterior, however, and the exterior is the region we study.
We therefore take the norms strictly outside the apparent horizon, over
$r_{\rm AH}+2\le r\le150$, where $r_{\rm AH}$ is the largest of the
coordinate radii of the horizon on the three grids.

We average the orders over the outputs from $t=4$ to $t=28$ and list
them in the last row of Table~\ref{tab:conv-scatter}. They lie between
$2.24$ and $2.62$, so every evolved field, $Q$ and the constraint
converge at second order.
For the constraint, Table~\ref{tab:conv-ratio} in
Appendix~\ref{sec:app-conv} lists the ratios of its norms on successive
grids at selected times. For exact second order these ratios equal $4$.
On the coarser pair of grids they stay close to $4$. On the finer pair
they are close to $4$ up to $t\simeq8$ and then rise to $\simeq6$, so
on this pair the constraint violation decreases faster than second
order. This
happens throughout the window, so it is not caused by a localised
feature.

The areal radius of the apparent horizon also converges at second
order. At $t=28$, the differences between its values on successive grids
shrink by a factor $4.3$. Assuming second order, its Richardson limit is
$\ThAH\simeq12.70$.

Finally, we check that the ADM mass is conserved. We read it at
$\Theta=250$, where $m(\Theta)$ is flat (Sec.~\ref{sec:method}). The
outer edge of the fixed grid falls inward, and this sphere leaves the
grid at $t\simeq20$. We therefore read the mass on the moving mesh at
the finest resolution ($N_r=51201$), where the outer edge stays at
$\Theta\simeq284$. Between $t=2$ and $t=28$, $m$ on this sphere grows by $0.07\%$. This is
neither a truncation error nor an effect of the outer boundary: at $t=28$
the outer edge of the fixed grid is at $\Theta\simeq235$, and on that
sphere the three fixed grids agree to $10^{-4}$ and the fixed grid and
the moving mesh agree to $2\times10^{-5}$. Nor is it a change of
$\MADM$, the limit of $m$ at large $\Theta$: on the spheres
$\Theta=100$--$250$ the change falls off as $\Theta^{-3}$ (it is $1.2\%$
at $\Theta=100$), and its extrapolation to $\Theta\to\infty$ is
about $0.01\%$.

\section{The mass function and the horizon}
\label{sec:mass}

\begin{figure}[htbp]
\centering
\maybegraphic[width=\columnwidth]{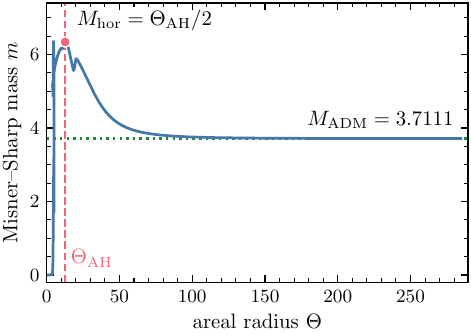}
\caption{Misner--Sharp mass $m$ (blue) against the areal radius $\Theta$
at $t=28$ in the baseline collapse. Outside the apparent horizon (red,
dashed) $m$ decreases outward from $\Mhor=\ThAH/2$ (red dot) to
$\MADM=3.7111$ (green, dotted). The narrow spike at $\Theta\approx5$,
inside the apparent horizon in the frozen interior, is a discretisation
artefact.}
\label{fig:mass}
\end{figure}

Figure~\ref{fig:mass} shows the Misner--Sharp mass at $t=28$, when the
apparent horizon is largest. Outside the horizon, $m$ decreases outward
from $\Mhor=\ThAH/2=6.35$ (Richardson limit of $\ThAH$,
Sec.~\ref{sec:conv})
to $\MADM=3.7111$ at large radius,
monotonically apart from a small feature just outside the horizon. The
decrease is slow, which is why $\MADM$ must be read at
$\Theta\gtrsim250$ (Sec.~\ref{sec:method}). The exterior therefore
carries a net energy $\MADM-\Mhor=-2.64$.

In spherical symmetry the mass function obeys~\cite{Hayward:1994bu}
\begin{equation}
  \begin{gathered}
  \partial_\Theta m|_t=\frac{\Theta^2}{2}\Bigl[\rho+\frac{A}{B}\,j\Bigr],\\
  A\equiv u^a\partial_a\Theta=\Theta\Bigl(\frac{Q}{2}-\frac{\trK}{3}\Bigr),
  \qquad B\equiv\partial_r\Theta ,
  \end{gathered}
  \label{eq:dmdTheta}
\end{equation}
where $\rho=T_{ab}u^au^b$ and $j=-T_{ab}u^ae^b$ are the energy density and
radial energy current measured by the \ae ther, $e^a$ is the unit radial
vector and $T_{ab}$ is the total (scalar plus \ae ther) stress
tensor.
The expansions of Eq.~\eqref{eq:expansions} are $\theta_\pm=B\pm A$, so
a sphere is untrapped ($\theta_\pm>0$) when $B>|A|$. The dominant energy
condition requires the energy-flux vector
$-T^a{}_bu^b=\rho\,u^a+j\,e^a$ to be causal and future-directed, i.e.\
$\rho\ge|j|$. The bracket in Eq.~\eqref{eq:dmdTheta} is then
non-negative on untrapped spheres: outside the apparent horizon, $m$ can
decrease outward only where the dominant energy condition is
violated~\cite{Hayward:1994bu}.

The outward decrease of $m$ in Fig.~\ref{fig:mass} therefore requires a
violation of the dominant energy condition outside the horizon. Equation~\eqref{eq:dmdTheta}
is linear in $T_{ab}$ and the scalar obeys the condition, so the
violation comes from the \ae ther, whose contribution to the exterior
energy is at most $-2.64$. Indeed, the \ae ther energy density is
negative everywhere outside the horizon except across the outgoing scalar
shell ($18\lesssim\Theta\lesssim20$ at $t=28$), and decays outward in the
long negative tail noted in Sec.~\ref{sec:method}.

In GR with matter obeying the dominant energy condition, integrating
Eq.~\eqref{eq:dmdTheta} from the horizon to infinity gives the
spherical Penrose inequality~\cite{Penrose:1973um,Malec:1994sy,Hayward:1994bu},
$\Mhor\le\MADM$. Here instead
\begin{equation}
  \frac{\Mhor}{\MADM}=1.71 ,
  \label{eq:mratio}
\end{equation}
a horizon more massive than the spacetime that contains it. Similarly,
the area theorem~\cite{Hawking:1971tu,Hawking:1973uf}, in its form for
apparent horizons~\cite{Hayward:1993wb,Ashtekar:2003hk}, would forbid
$\ThAH$ from decreasing, whereas in our run $\ThAH$ peaks at $t\simeq28$
and then decreases slowly, from $12.6$ to $11.8$ by $t=150$ on the grid
of Fig.~\ref{fig:mesh}(b). The area theorem relies only on the null
energy condition, which the \ae ther stress tensor does not satisfy in
general~\cite{Akhoury:2016mrw}.

\section{The spin-0 mode}
\label{sec:spin0}

\subsection{Principal part and characteristic variables}
\label{sec:principal}

With $\partial_r\psi$ as an extra variable, Eqs.~\eqref{eq:evolution}
are first order in time and space, and the eigenvalues and eigenvectors
of their principal part give the characteristic speeds and variables.
The equations for $\psi$, $P$ and $\Theta$ contain no derivatives of
$\trK$ or $a_r$, so the principal symbol is block-triangular and the
spin-0 speeds are those of the $(\trK,a_r)$ block, whose principal part
is
\begin{equation}
  \partial_t\mathbf u = \mathsf M\,\partial_r\mathbf u+\dots,\qquad
  \mathsf M=\begin{pmatrix}\beta & N a_N\\ N b_N & \beta\end{pmatrix},
  \label{eq:principal}
\end{equation}
with $\mathbf u=(\trK,a_r)^{T}$ and
\begin{equation}
  \begin{gathered}
  a_N=\frac{c_{14}-2}{2+c_{13}+3c_2},\qquad
  b_N=-\frac{c_{123}}{c_{14}(1-c_{13})},\\
  a_Nb_N=v_0^2 .
  \end{gathered}
  \label{eq:aNbN}
\end{equation}
The diagonal entries describe advection with the \ae ther observers,
$\dot r=-\beta$; the off-diagonal ones couple the two fields as in a
first-order wave equation, whose squared speed is that of the linear
spin-0 mode, Eq.~\eqref{eq:speeds}. The eigenvalues
$\lambda_\pm=\beta\pm Nv_0$ give characteristics $\dot r=-\beta\mp Nv_0$,
i.e.\ speeds $\pm v_0$ relative to the \ae ther in its proper time and
proper radial distance ($\gamma_{rr}=1$), whatever the lapse and shift.
Franchini, Herrero-Valea and Barausse~\cite{Franchini:2021bpt} found the
same speeds for the infrared limit of Ho\v{r}ava gravity in spherical
symmetry, with the same slicing but zero shift and $\gamma_{rr}\neq1$,
using derivatives of the metric functions as first-order variables:
their spin-0 characteristics, $\dot r=\pm v_0N/\sqrt{\gamma_{rr}}$,
again correspond to a speed $v_0$ relative to the
\ae ther~\cite{Jacobson:2010mx}.

The left eigenvectors of $\mathsf M$ give the characteristic variables
\begin{equation}
  \wout=-\frac{b_N}{v_0}\,\trK+a_r,\qquad
  \win=+\frac{b_N}{v_0}\,\trK+a_r ,
  \label{eq:wout}
\end{equation}
which obey $\partial_t w-\lambda\,\partial_r w=(\text{lower order})$,
with $\lambda=\lambda_-$ for $\wout$, outgoing relative to the \ae ther,
and $\lambda=\lambda_+$ for $\win$. They become exact characteristic
variables of the full system far from the source, where the terms
coupling them to $\Theta$ and $\psi$, proportional to $a_r/\Theta$ and
$P$, fall off.
The outgoing and ingoing scalar combinations are
$P\mp\partial_r\psi$, which move along $\dot r=-\beta\pm N$.

\subsection{The spin-0 wave far from the source}
\label{sec:infinity}

Far from the source the spacetime is nearly flat and the fields are
small, so the spin-0 equations can be solved exactly. We solve them in
the static frame of flat space, $N=1$ and zero shift, with the areal
radius $\Theta=r$ as radial coordinate. Our gauge differs from this frame
only by a rigid translation of the radial labels on each slice,
$r\to r+c(t)$, which keeps $\gamma_{rr}=1$ and gives the shift
$\beta=\dot c$. Since $\beta\simeq-1$ far out, $c(t)\simeq-t$: a point at
fixed $r$ falls inward in areal radius at about unit speed
(Sec.~\ref{sec:method}), and a time derivative at fixed areal radius
becomes $\partial_t-\beta\partial_r$ in our coordinates. Keeping only the
terms linear in the fields, the $\trK$ and $a_r$ equations of
Eq.~\eqref{eq:evolution} become
$\partial_t\trK=a_N\,r^{-2}\partial_r(r^2a_r)$ and
$\partial_t a_r=b_N\,\partial_r\trK$; the scalar drops out, since it
enters them only quadratically. Together they give the wave equation
$\partial_t^2\trK=v_0^2\,r^{-2}\partial_r(r^2\partial_r\trK)$, and an
outgoing pulse is
\begin{equation}
  \begin{gathered}
  \trK=\frac{f(u)}{r},\qquad
  a_r=-\frac{b_N}{v_0}\,\frac{f(u)}{r}-b_N\,\frac{F(u)}{r^2},\\
  u=t-\frac{r}{v_0},\quad F'=f,
  \end{gathered}
  \label{eq:asymptotic}
\end{equation}
where the profile $f$ is set by the source and $F$ is its integral.
Besides the $1/r$ falloff of any spherical wave, $a_r$ has a $1/r^2$
part, which builds up as the pulse goes by and remains behind it. In
terms of the characteristic variables of Eq.~\eqref{eq:wout},
\begin{equation}
  \wout=-\frac{2b_N}{v_0}\,\frac{f(u)}{r}-b_N\,\frac{F(u)}{r^2},
  \qquad
  \win=-b_N\,\frac{F(u)}{r^2}.
  \label{eq:wout-asym}
\end{equation}
The radiation is carried by $\wout$, and $r\,\wout\to-2b_Nf(u)/v_0$ is
the spin-0 waveform we extract. The ingoing variable $\win$ has no $1/r$
part, but it is not zero: it changes while the pulse goes by. This
matters for the outer boundary condition below.

 At $r=0$ we impose regularity: $\psi$, $P$,
$\trK$ and $Q$ are even in $r$, and $\Theta$, $a_r$ and $\beta$ are odd.
At the outer boundary a condition is needed only for the fields that
enter the grid there, $\win$ for the spin-0 pair and $P+\partial_r\psi$
for the scalar; the outgoing fields simply leave. Physically, no
radiation should come in from infinity. At finite radius, however, a
purely outgoing wave has a nonzero entering field, one power of $r$
below the radiation: $\win=-b_NF(u)/r^2$ by Eq.~\eqref{eq:wout-asym}, and
$P+\partial_r\psi=-g(u)/r^2$ for a scalar wave $\psi=g(u)/r$. Setting
these fields to zero at $r_{\max}$, their value at infinity, therefore
reflects part of the wave, by the ratio of the two falloffs, i.e.\ about
the pulse width over $r_{\max}$. Instead we impose non-reflecting conditions,
which require the fields at $r_{\max}$ to behave as for an outgoing wave: $r\trK=f(u)$ and $r\psi=g(u)$ must be constant along
the outgoing characteristics, of speed $v_0$ and $1$. For the spin-0
pair this is equivalent to $\partial_t\win=-b_N\trK/r$, the form we
implement~\cite{Sarbach:2012pr}; on the collapse background, with the
time derivative at fixed areal radius divided by the lapse and $\Theta$
for $r$, it reads
\begin{equation}
  (\partial_t-\beta\partial_r)\win=-Nb_N\,\frac{\trK}{\Theta}.
  \label{eq:bc-win}
\end{equation}
For the scalar we impose that $\Theta\psi$ be constant along the outgoing
light rays; for a linear wave on flat space this is equivalent to
$\partial_t(P+\partial_r\psi)=-P/r$, the analogue of
Eq.~\eqref{eq:bc-win}, and the entering field again changes at a rate set
by its $1/r^2$ part.

These conditions reflect very little. A spin-0 pulse
$\trK\propto\exp[-(r-50)^2/16]$ crossing a flat-space boundary at
$r_{\max}=100$ and $150$, for $v_0=1$ and $1.267$, is reflected into
$\win$ below $10^{-5}$ of the incident $\wout$, against $3.5\%$
($r_{\max}=100$) and $2.3\%$ ($r_{\max}=150$) when $\win$ is set to zero.
A scalar pulse is reflected with $0.03\%$ of its amplitude.

\subsection{The spin-0 waveform and its speed}
\label{sec:results}
We analyse two collapses of the same initial shell: one with the
baseline couplings, for which $v_0=1$ and the spin-0 and scalar
signals travel together, and one with $c_2=0$, for which $v_0=1.267$ and
the two signals separate (Sec.~\ref{sec:method}).
We first extract the spin-0 waveform from the outgoing characteristic
variable $\wout$ and check that it falls off as $1/\Theta$. We then
measure the speed of the signal in two independent ways: with a linear
probe pulse on the collapse background, and with the burst in the
Misner--Sharp luminosity, used as a clock between spheres.

\begin{figure*}[t]
\centering
\maybegraphic[width=\textwidth]{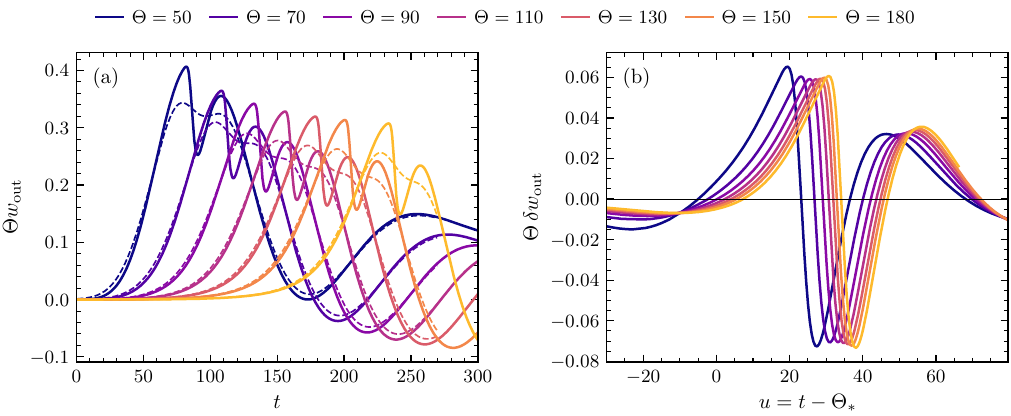}
\caption{Outgoing spin-0 characteristic $\wout$ in the
baseline collapse ($v_0=1$), on spheres of fixed areal radius. (a) $\Theta\wout$ against $t$ (solid) and its background $\Theta\bar w_{\rm out}$
(dashed), a Gaussian running mean of $\wout(t)$ of width
$\sigma=10$ at fixed $\Theta$. (b) The pulse
$\Theta\,\delta\wout=\Theta(\wout-\bar w_{\rm out})$ against retarded time
$u=t-\Theta_*$. Its amplitude is the same on every sphere, as expected
from the $1/\Theta$ falloff of Eq.~\eqref{eq:wout-asym}.}
\label{fig:wout}
\end{figure*}

At fixed $\Theta$, $\Theta\wout(t)$ shows a broad outgoing hump followed
by a narrow pulse (Fig.~\ref{fig:wout}(a)). To isolate the pulse, we split $\wout$ on each sphere into a slowly
varying background $\bar w_{\rm out}$ and a fast remainder
$\delta\wout$, smoothing $\wout$ in time with a Gaussian kernel:
\begin{equation}
  \bar w_{\rm out}(\Theta,t)=
  \frac{\int\wout(\Theta,t')\,e^{-(t-t')^2/2\sigma^2}\,dt'}
       {\int e^{-(t-t')^2/2\sigma^2}\,dt'},
  \label{eq:wbar}
\end{equation}
and $\delta\wout=\wout-\bar w_{\rm out}$.
With a width $\sigma=10$, the remainder isolates the
pulse $\Theta\,\delta\wout$, whose amplitude depends on $\sigma$ but is
the same on every sphere for any $\sigma\in[5,15]$
(Fig.~\ref{fig:wout}(b)), as expected from Eq.~\eqref{eq:wout-asym}.
To compare spheres we use the retarded time
\begin{equation}
  u=t-\frac{\Theta_*}{v_0},\qquad
  \Theta_*=\Theta+2\MADM\ln\!\left(\frac{\Theta}{2\MADM}-1\right),
  \label{eq:retarded}
\end{equation}
the flat-space $u$ of Eq.~\eqref{eq:asymptotic} with $r$ replaced by
the Schwarzschild tortoise coordinate $\Theta_*$, which accounts
approximately for the slowing of the wave near the black hole.
The narrow pulse drifts slowly in $u$ (Fig.~\ref{fig:wout}(b)), which we attribute to this
approximation.

\begin{figure}[t]
\centering
\maybegraphic[width=\columnwidth]{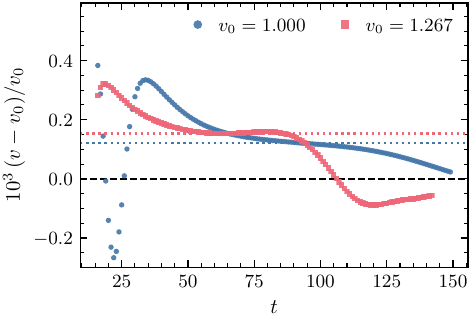}
\caption{Relative deviation $(v-v_0)/v_0$ of the speed
$v=(\dot r_{\rm pk}+\beta)/N$ of a linear spin-0 probe, measured relative
to the \ae ther on the background of the collapse, against $t$, for
$v_0=1$ (blue circles) and $v_0=1.267$ (red squares); $r_{\rm pk}$ is the
coordinate radius of the peak of the probe. The dashed line is $v=v_0$
and the dotted lines are the medians of the measured speeds, which agree
with $v_0$ to $0.01\%$ and $0.02\%$, respectively.}
\label{fig:probe}
\end{figure}

To test the spin-0 propagation, we add to the initial data a linear probe pulse of $\trK$ (amplitude
$10^{-6}$, centred at $r=150$), isolate it as the difference with the run
without it, and measure the speed of the peak of its $\wout$ relative to
the \ae ther, $v=(\dot r_{\rm pk}+\beta)/N$.
As shown in Sec.~\ref{sec:principal}, the outgoing spin-0
characteristic moves at exactly $v_0$ relative to the \ae ther, measured in its
proper time and proper radial distance, so
this measurement requires no approximation of the background. Figure~\ref{fig:probe} shows $(v-v_0)/v_0$ during the
evolution. The scatter at early times, $t\lesssim40$, comes from the
splitting of the initial $\trK$ pulse, which has both $\wout$ and
$\win$ components, into its outgoing and ingoing parts; afterwards, the
deviation stays below $3\times10^{-4}$. The median speed agrees with $v_0$ to $0.01\%$ for $v_0=1$ and to $0.02\%$ for $v_0=1.267$.

\begin{figure*}[t]
\centering
\maybegraphic[width=\textwidth]{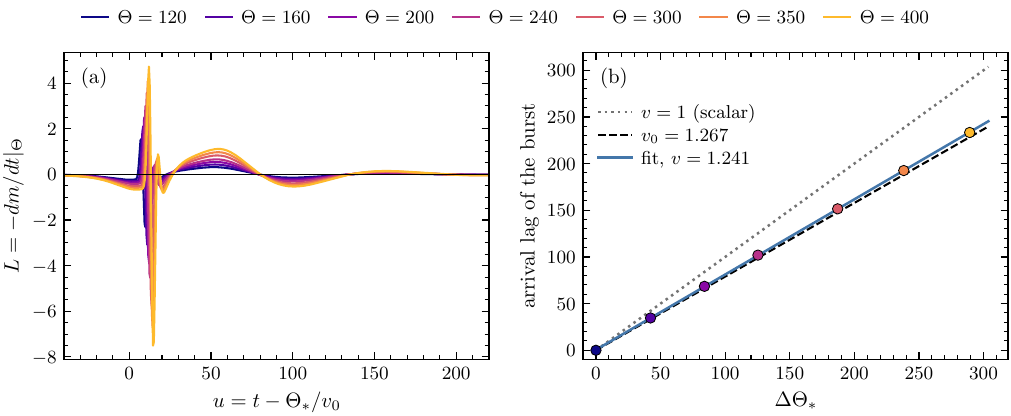}
\caption{(a) Misner--Sharp luminosity $L=-dm/dt|_\Theta$ through
spheres of constant $\Theta$, against retarded time $u=t-\Theta_*/v_0$,
for $v_0=1.267$. It is dominated by the \ae ther Coulomb term
(Sec.~\ref{sec:flux}). (b) The burst as a clock: its arrival lag at
each sphere (dots, coloured as in (a)) against the tortoise distance
$\Delta\Theta_*$ from the innermost sphere. A linear fit (blue) gives the
speed $1.241$, $2\%$ below $v_0=1.267$ (black, dashed), which we
attribute to the Schwarzschild tortoise approximation. The dotted line
is the speed of the scalar, $v=1$.}
\label{fig:lum}
\end{figure*}

To carry out an independent measurement of the
speed of the nonlinear signal emitted by the collapse, without
using the characteristic variables, we use the Misner--Sharp
luminosity through a sphere of fixed areal radius,
$L=-dm/dt|_\Theta$ (its decomposition into \ae ther and scalar fluxes is
given in Sec.~\ref{sec:flux}).
For $v_0=1.267$, $L$ shows on every sphere a sharp burst at
$u\simeq10$--$15$, followed by a broad oscillation (Fig.~\ref{fig:lum}(a));
the burst coincides with the narrow pulse of $\wout$. We take the
arrival time of the burst on each of the spheres $\Theta=120$--$400$
and plot its lag with respect to the innermost sphere against
the tortoise distance $\Delta\Theta_*$ between the spheres
(Fig.~\ref{fig:lum}(b)). A signal moving at speed $v$ in tortoise
distance gives a straight line of slope $1/v$. The arrival times lie on
a line, and a linear fit gives $v=1.241$, $2\%$ below $v_0=1.267$, which
we attribute to the Schwarzschild tortoise approximation in
Eq.~\eqref{eq:retarded}.
The burst is therefore
carried by the spin-0 mode, not by the scalar field, which propagates at the speed of light.

\subsection{Mass conservation and radiated energy}
\label{sec:flux}
We now turn to the energy carried by the wave. We first
check that, at every time, the change of the Misner--Sharp mass on each
sphere is accounted for by the flux of the stress-energy tensor
through it. We then show that most of this flux is
the \ae ther Coulomb field, which is linear in the wave amplitude, and
remove it to obtain the radiated energy.

\begin{figure*}[t]
\centering
\maybegraphic[width=\textwidth]{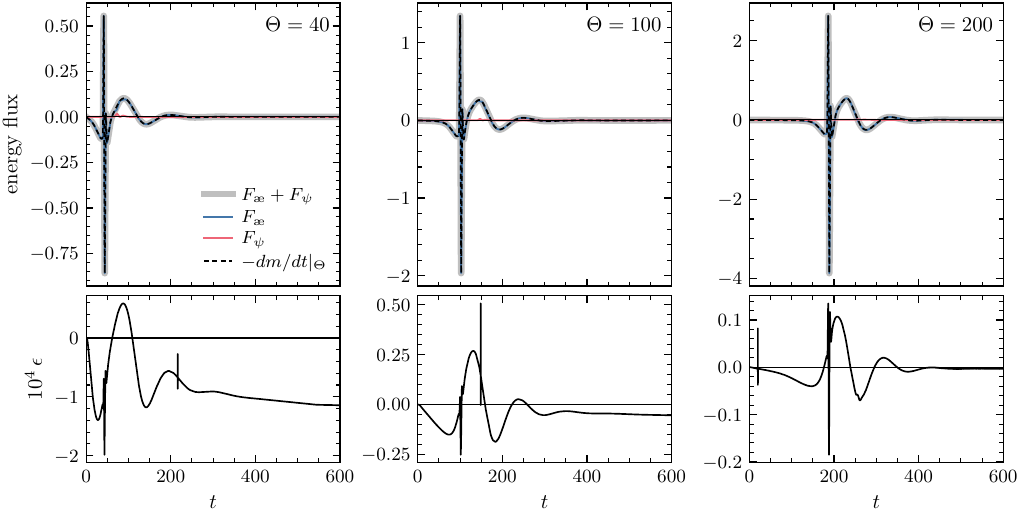}
\caption{Test of the flux identity, Eq.~\eqref{eq:lum}, for $v_0=1.267$
on the spheres $\Theta=40$, $100$ and $200$ (columns). Top row: energy
flux through the sphere against $t$: $F_{\text{\ae}}$ (blue), computed
from the \ae ther stress tensor, $F_\psi$ (red), their sum
$F_{\text{\ae}}+F_\psi$ (thick grey) and $-dm/dt|_\Theta$ (black,
dashed). Bottom row: the relative residual
$\epsilon=(F_{\text{\ae}}+F_\psi+dm/dt|_\Theta)/\max|dm/dt|_\Theta|$, in
units of $10^{-4}$, which decreases outward.}
\label{fig:fluxid}
\end{figure*}

The time derivative of the Misner--Sharp mass at fixed areal radius equals the flux of the stress-energy tensor through the sphere~\cite{Hayward:1994bu}. Using the field equations $G_{ab}=T_{ab}$, one obtains the luminosity $L$, split into \ae ther and scalar-field contributions as
\begin{equation}
  \begin{gathered}
  L\equiv-\left.\frac{dm}{dt}\right|_\Theta=F_{\text{\ae}}+F_\psi ,\\
  F_X=\frac{N\Theta^2}{2}\left[A\,(\rho_X+p_X)+\frac{A^2+B^2}{B}\,j_X\right],
  \end{gathered}
  \label{eq:lum}
\end{equation}
with $F>0$ outward, $A$, $B$, $\rho_X$ and $j_X$ as in
Sec.~\ref{sec:mass} for each component $X$ of $T_{ab}$, and
$p_X=T_{ab}e^ae^b$ the radial pressure measured by the
\ae ther.
For $v_0=1.267$,
the two sides of Eq.~\eqref{eq:lum} agree to within $0.02\%$ of the
peak flux at all times on every sphere from $\Theta=40$ to $200$
(Fig.~\ref{fig:fluxid}).
The scalar flux is small, $|F_\psi|\lesssim0.02$, against \ae ther-flux
peaks of $0.9$--$4$, mostly the Coulomb contribution discussed below.

\begin{figure*}[t]
\centering
\maybegraphic[width=\textwidth]{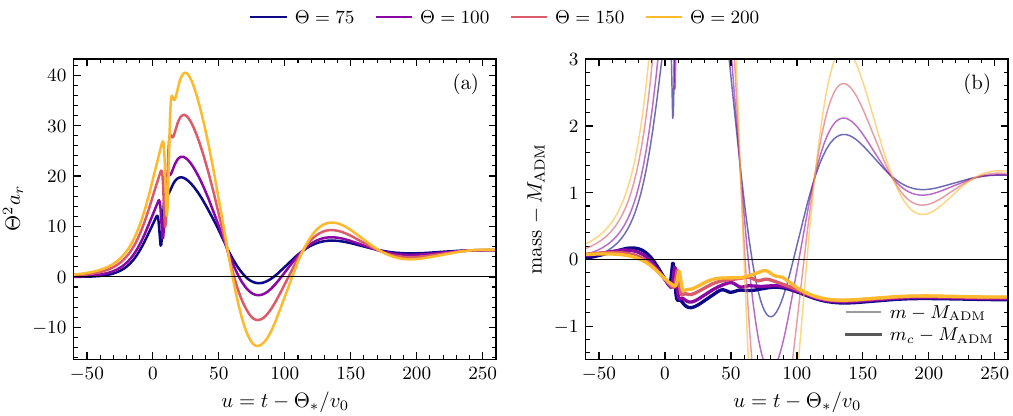}
\caption{The \ae ther Coulomb field left behind by the spin-0 pulse
($v_0=1.267$), on spheres with constant $\Theta$ against
retarded time $u=t-\Theta_*/v_0$. (a) $\Theta^2a_r$;
during the pulse its amplitude
grows with $\Theta$ (roughly as $\Theta^{0.7}$).
(b) $m-\MADM$ (thin lines) and $m_c-\MADM$ (thick lines), with $m_c$ defined in
Eq.~\eqref{eq:mc}. Removing the Coulomb term reduces the swing and $m_c$ settles below $\MADM$
at all radii.}
\label{fig:coulomb}
\end{figure*}

However, the luminosity gives a confusing answer
for the radiated energy. As the pulse crosses a sphere, $m$ swings by an
amount that grows with radius, and behind
the pulse it settles above $\MADM$
(Fig.~\ref{fig:coulomb}(b), thin lines), so that $\int L\,du$ is
negative on every sphere. We explain this behaviour below.

Let us first note that the spin-0 front sets up the acceleration $a_r\propto\Theta^{-2}$ of the
\ae ther around a black hole, absent from the initial data: behind the pulse $\Theta^2a_r$ relaxes to the same value on
every sphere
(Fig.~\ref{fig:coulomb}(a)), the $F(u)/r^2$ term of
Eq.~\eqref{eq:asymptotic}. 
We call such a contribution the \ae ther Coulomb field.

At large radius, where $\trK,Q\to0$, the
Hamiltonian constraint reduces to 
\begin{equation}
  \begin{split}
  \partial_\Theta m={}&\frac{c_{14}}{2}\,\partial_r\Theta\;
  \partial_\Theta\!\left(\Theta^2a_r\right)\\
  &+\bigl(\text{terms quadratic in }a_r,\,\trK,\,Q,\,\psi\bigr),
  \end{split}
  \label{eq:dm-coulomb}
\end{equation}
with $\partial_r\Theta=1+O(M/\Theta)$. To leading order the linear term
is a total derivative, so during the pulse, when the amplitude of
$\Theta^2a_r$ grows with $\Theta$, the swing of $m$ grows with radius
and dominates the luminosity
(Fig.~\ref{fig:lum}(a)). Being linear in the amplitude, it is not
radiated energy, and we remove it, defining the corrected mass and the cumulative radiated energy as:
\begin{equation}
  m_c\equiv m-\frac{c_{14}}{2}\,\Theta^2a_r ,
  \qquad
  E(\Theta,u)\equiv\MADM-m_c(\Theta,u),
  \label{eq:mc}
\end{equation}
where $E$ is the energy that has crossed the sphere $\Theta$ up to
retarded time $u$. Behind the pulse $m_c$ settles below $\MADM$ on every
sphere, Fig.~\ref{fig:coulomb}(b). This discussion agrees with the definition of the quasi-local version of the total energy of an asymptotically flat Einstein--\ae ther
spacetime~\cite{Eling:2005zq,Garfinkle:2011iw},
\begin{equation}
  M_{\text{\ae}}=\MADM-\frac{c_{14}}{8\pi G}\oint_\infty r^a a_a\,dA ,
  \label{eq:Mae}
\end{equation}
which gives $\MADM-\tfrac{c_{14}}{2}\lim_{\Theta\to\infty}\Theta^2a_r$ in
spherical symmetry, positive for a class of solutions when $c_{14}<2$~\cite{Garfinkle:2011iw}.

For the superluminal case $c_2=0$ with $\MADM=3.708$,
$E(\Theta)$ differs from the energy radiated to infinity by corrections of order $M/\Theta$, because the spheres lie at finite radius. We therefore extrapolate to infinity with the two-parameter fit
$E_\infty+b/\Theta+O(1/\Theta^2)$ over $\Theta=75$--$200$, and the scalar flux gives $E_\psi=0.197$.
We obtain
\begin{equation}
  \begin{gathered}
  E_\infty=0.532\pm0.023=(0.144\pm0.006)\,\MADM,\\
  E_{\text{\ae}}=E_\infty-E_\psi=0.335=0.090\,\MADM,
  \end{gathered}
  \label{eq:Einf}
\end{equation}
so the spin-0 wave carries off $9\%$ of the ADM mass,
a positive energy as required for the linear spin-0 mode when
$c_{14}(2-c_{14})>0$~\cite{Eling:2005zq,Jacobson:2008aj}.

\begin{figure}[t]
\centering
\maybegraphic[width=\columnwidth]{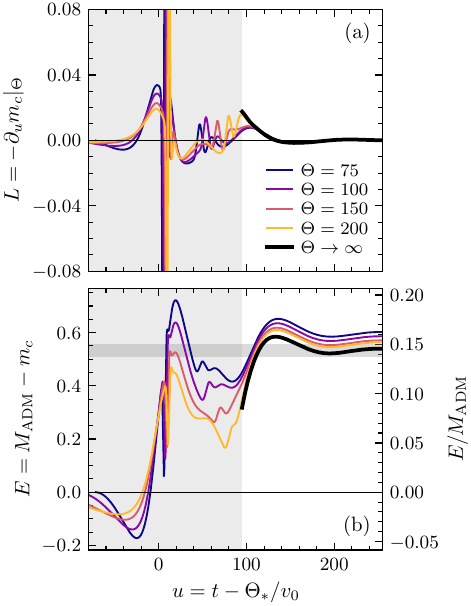}
\caption{Energy carried by the spin-0 wave ($v_0=1.267$), on spheres
of constant $\Theta$ against retarded time $u=t-\Theta_*/v_0$. (a) Rate
$L=-\partial_um_c|_\Theta$ and its extrapolation $L_\infty$ to infinity
(black). (b) Energy $E=\MADM-m_c$ that has crossed each sphere (left
axis; $E/\MADM$ on the right axis) and its extrapolation $E_\infty(u)$
from a fit $E_\infty+b/\Theta$ (black). The extrapolations are shown only
after the pulse has crossed the spheres, outside the shaded interval.
The horizontal grey band is the value $E_\infty=0.53\pm0.02$ of
Eq.~\eqref{eq:Einf}.}
\label{fig:fluxinf}
\end{figure}

Figure~\ref{fig:fluxinf} follows the corrected mass in
time. Once the Coulomb term is removed, the rate
$L=-\partial_u m_c|_\Theta$ (panel (a)) is more than an order of magnitude
smaller than the uncorrected luminosity of Fig.~\ref{fig:lum}(a), and
the cumulative energy $E$ (panel (b)) is nearly the same on all
spheres once the pulse has passed; its extrapolation to infinity, shown once the pulse has
crossed all spheres, settles on the value of Eq.~\eqref{eq:Einf}.
Before the front arrives,
the freely falling \ae ther makes $E$ slightly negative; $E$ then overshoots its asymptotic value before settling to the total radiated energy. Integrating Eq.~\eqref{eq:dm-coulomb} inward from a sphere not yet
reached by the wave shows that, up to $O(M/\Theta)$ corrections, $E$ is
the integral of the quadratic and higher-order field energy outside
$\Theta$. The transient overshoot is a consequence of the fact that this quadratic energy density is not positive definite, due to shear contributions from the \ae ther stress tensor, yielding $-\tfrac{3}{4}c_{13}Q^2$, which is negative for $c_{13}>0$.

\subsection{The spin-0 mode in the post-Newtonian expansion}
\label{sec:pn_comparison}

Ref.~\cite{Taherasghari:2023rwn} derived the Einstein--\ae ther field
equations to 2.5PN order, and Ref.~\cite{Taherasghari:2025mlf} found
that the resulting energy flux does not depend on the spin-0
propagation speed $v_L$; this was interpreted as evidence
that the spin-0 mode is non-radiative. Here we point out a subtlety in
the interpretation of the truncated field equations: a non-local field
redefinition involving inverse Laplacians can shift the PN order at
which the propagation speed appears explicitly in a given field
equation. A single-field example of this mechanism is given in
Sec.~\ref{app:pn_toy}.

In this subsection and in Appendix~\ref{app:pn_redefinition} we adopt the notation of Ref.~\cite{Taherasghari:2023rwn}. In
particular, $N$, $B$, $K^j$, $K^j_{\text{\ae}}$ and $R$ denote the PN
potentials defined there, and should not be confused with the lapse,
$\partial_r\Theta$ and $\trK$ used above. As in
Refs.~\cite{Taherasghari:2023rwn,Taherasghari:2025mlf}, where
$c_3=-c_1$ is assumed because of the GW170817 bound, the PN analysis is
carried out for $c_{13}=0$, for which the spin-0 speed $v_0$ of
Eq.~\eqref{eq:speeds} reduces to the speed $v_L$ of Ref.~\cite{Taherasghari:2023rwn},
$v_L^2=c_2(2-c_{14})/[c_{14}(2+3c_2)]$.

In the spherically symmetric scalar sector of the PN-expanded
Einstein--\ae ther equations of Ref.~\cite{Taherasghari:2023rwn}, eliminating the
auxiliary potentials yields an equation for the lapse perturbation that contains
an explicit $v_L$-dependent term at 1PN order. Truncating the field
redefinition of Ref.~\cite{Taherasghari:2023rwn} at first order in the PN parameter
removes this term, but the dependence on $v_L$ reappears at the next
order: the first surviving $v_L$-dependent term in the transformed
lapse equation is 2PN,
\begin{equation}
\left(1-\frac{c_{14}}{2}\right)\Box N
-\kappa_1\eta^2\partial_t^4\nabla^{-2}N+\cdots=0,
\label{eq:pn_first_surviving}
\end{equation}
where $\Box\equiv\nabla^2-\eta\partial_t^2$, $\eta$ counts PN
orders, and the coefficient $\kappa_1\propto v_L^{-4}$ is given in
Eq.~\eqref{eq:kappa1}. The derivation is given in
Appendix~\ref{app:pn_redefinition}.

The reason why $v_L$ reappears becomes transparent on a linear mode
$e^{-i\omega t}f(r)$ with $\nabla^2f=-k^2f$. Defining
$x\equiv\eta\omega^2/k^2$ and $y\equiv x/v_L^2$, the reduced lapse
operator before the redefinition, $s_{-1}(x)$ [Eq.~\eqref{eq:sminus1}],
and the diagonal lapse operator of Eq.~(4.13a) of Ref.~\cite{Taherasghari:2023rwn},
represented on this mode by $-(1-\tfrac12c_{14})(1-x)$, are related
exactly by
\begin{equation}
\begin{split}
\Gamma(x)&=\frac{-(1-\frac12c_{14})(1-x)}{s_{-1}(x)}\\
&=\frac{1-x/x_*}{1-x/v_L^2}=1+\frac{c_{14}}2\frac{y}{1-y},
\end{split}
\label{eq:G_exact}
\end{equation}
where $x_*\equiv2v_L^2/(2-c_{14})$. The factor containing the scalar
propagation speed is an inverse propagator,
\begin{equation}
\frac{y}{1-y}
=\frac{\eta\partial_t^2}{v_L^2}
\left(\nabla^2-\frac{\eta}{v_L^2}\partial_t^2\right)^{-1}.
\label{eq:inverse_propagator}
\end{equation}
Removing the $v_L$ dependence from the lapse equation to all orders
would therefore require the whole PN series of this inverse
propagator, each term of which carries one more PN order and one more
inverse Laplacian. The first-order redefinition retains only the first
term, and the first neglected one reappears at 2PN as the $\kappa_1$
term in Eq.~\eqref{eq:pn_first_surviving}. Moreover, $\Gamma(x)$ is singular
at the physical scalar pole $x=v_L^2$: its cancellation of the pole in
the reduced lapse operator must not be interpreted as the removal of a
physical degree of freedom by an admissible, globally invertible field
redefinition.

For the first-order truncation of Eq.~(4.7) of Ref.~\cite{Taherasghari:2023rwn}, the
explicit dependence of the lapse equation on $v_L$ is thus removed at
1PN and reappears at 2PN. We expect the same mechanism to operate when
the full redefinition (4.7) is used, as in Ref.~\cite{Taherasghari:2023rwn}: each
additional order retained in the transformation reproduces one more
term of the PN series of the inverse propagator in
Eq.~\eqref{eq:inverse_propagator}, and therefore pushes the first
explicit appearance of $v_L$ to higher PN order, plausibly beyond the
2.5PN order at which the energy flux was computed in
Ref.~\cite{Taherasghari:2025mlf}. The absence of $v_L$
from the flux at that order would then be a consequence of the
non-local choice of variables, rather than evidence that the spin-0
mode is non-radiative. In other words, the disappearance of the
propagation speed from the field equations at a given PN order, after a
non-local field redefinition, does not imply that the corresponding
mode does not radiate.

This was one of the main motivations for the analysis presented in this
section, in which we tracked the spin-0 field directly in nonlinear
evolutions of spherical collapse, without truncating a PN expansion or
performing non-local field redefinitions. The two analyses cover
different regions of parameter space: the PN calculation above assumes
$c_{13}=0$, whereas our simulations have $c_{13}\neq0$. In particular,
the run used to measure the radiated energy has $c_2=0$, so that the
spin-0 mode propagates only because $c_{13}\neq0$. The comparison is
therefore not a quantitative one between energy fluxes; it concerns the
mechanism by which a propagating spin-0 mode can disappear from a
truncated PN expansion while, as our simulations show, still carries
energy to infinity.

\section{Large $c_1$: transient white holes}
\label{sec:wh}

Akhoury, Garfinkle and Gupta~\cite{Akhoury:2016mrw} found that for large
$c_1$ the collapse of a scalar shell can produce an anti-trapped
surface instead of a trapped one, the defining feature of a white hole.
They reported that the anti-trapped region is transient and that the fields
eventually disperse, but also that sharp features appear which their grids
could not fully resolve. We revisit this case at higher resolution, with the numerical scheme of Sec.~\ref{sec:method}.

We take $c_1=0.8$, $c_3=c_4=0$ and
$c_2=-c_1^3/(2-4c_1+3c_1^2)=-0.7111$, which keeps $v_0=1$. The shell is the one of Eq.~\eqref{eq:id},
made initially ingoing as in Ref.~\cite{Akhoury:2016mrw}:
\begin{equation}
  \begin{gathered}
  \psi(0,r)=\frac{A_0}{100}\,r^2\exp\!\Bigl[-\Bigl(\frac{r^2-r_0^2}{\sigma^2}\Bigr)^{2}\Bigr],\\
  P(0,r)=\partial_r\psi+\frac{\psi}{r},
  \end{gathered}
  \label{eq:wh-id}
\end{equation}
so that $r\psi$ initially moves inward at the speed of light with $r_0=10$ and $\sigma=4$. Besides $\theta_+$, we use the ingoing
expansion $\theta_-$ of Eq.~\eqref{eq:expansions}. Where
$\theta_-<0$ the sphere is anti-trapped: ingoing light rays diverge, as in a
white hole. In GR the energy conditions forbid anti-trapped surfaces from
forming out of regular data; the negative energy density of the \ae ther
(Sec.~\ref{sec:mass}) removes this obstruction.

Whether a white hole or a black hole forms depends on the amplitude. At
$A_0=0.12$ the data are initially untrapped everywhere
but $\theta_+$ crosses zero by $t\simeq2.5$ and a black hole forms.
With $A_0=0.10$, after a brief dip of $\theta_+$ below zero at
$t\simeq6$--$17$, an anti-trapped region appears at $t\simeq20$ and
grows, its outer edge moving out with the escaping pulse, until it
splits at $t\simeq36$. The inner part shrinks after $t\simeq40$ and
disappears by $t\simeq44$, while a thin anti-trapped layer stays around
the peak of the escaping pulse and travels outward with it
(Fig.~\ref{fig:wh}(a)). The layer is present on all our grids and
narrows with the pulse, from about $0.9$ at $t=40$ to $0.06$ at $t=80$
on the refined grids, which agree on its width to within $20\%$; its
depth, $\min\theta_-$, still grows with resolution.
 This
reproduces the transient white hole of Ref.~\cite{Akhoury:2016mrw}, whose
anti-trapped surface first appears at a similar time ($t\simeq17$); their
Fig.~11 also shows trapped surfaces at higher amplitude. The two amplitude
normalisations differ, so the thresholds cannot be compared directly.

\begin{figure*}[t]
\centering
\maybegraphic[width=\textwidth]{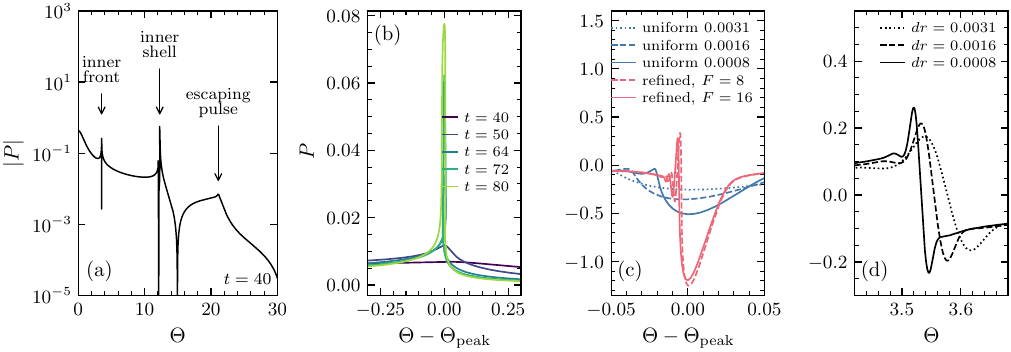}
\caption{The three sharp features of the white-hole run ($c_1=0.8$,
$A_0=0.10$), shown in $P$. (a) $|P|$ against $\Theta$ at $t=40$ (grey
line in Fig.~\ref{fig:wh}(a)) on the uniform grid with $dr=0.0008$. (b) The escaping pulse at $t=40$--$80$,
while its peak moves from $\Theta\simeq21$ to $62$, on the grid with a
region eight times finer that follows it, against the distance from the
peak. (c) The inner shell at $t=30$, against the distance from its peak,
on the three uniform grids (blue; $dr=0.0031$, $0.0016$, $0.0008$) and on
two grids refined around the shell (red; $F=8$ and $16$). (d) The inner
front at $t=40$ on the three uniform grids.}
\label{fig:sharp}
\end{figure*}

\begin{figure*}[t]
\centering
\maybegraphic[width=\textwidth]{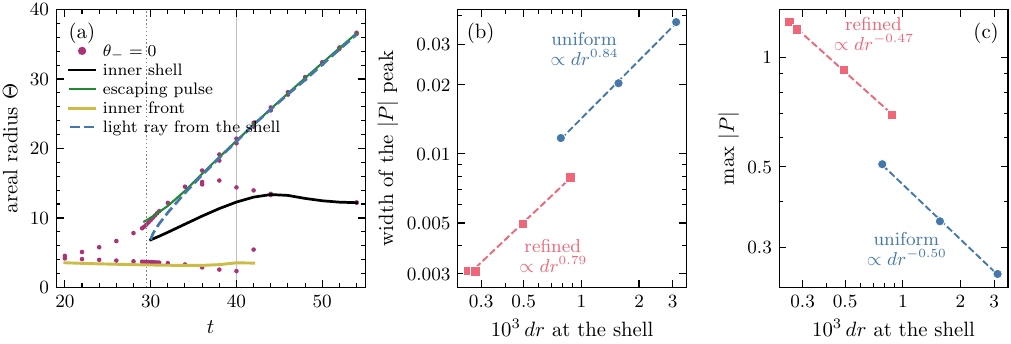}
\caption{The transient white hole at $c_1=0.8$, $A_0=0.10$. (a) Areal radius against $t$,
on the uniform grid with $dr=0.0008$, of the edges of the anti-trapped
region ($\theta_-=0$, purple dots), of the inner shell (peak of $|P|$,
black solid), of the escaping pulse (largest local maximum of $|P|$
outside the shell, green), of the inner front (steepest slope of $P$ at
$1.5<\Theta<5$, yellow) and of the outgoing light ray leaving the inner
shell at $t=30$ (blue dashed). The dotted vertical line marks
$t\simeq29.5$, when the inner shell steepens, and the grey line $t=40$,
the profile of Fig.~\ref{fig:sharp}(a). The ray joins the escaping
pulse by $t\simeq34$. The outermost branch of purple dots is a thin
anti-trapped layer, less than one unit wide (its two edges overlap at
this scale), that travels with the peak of the escaping pulse. (b) Width (full width at half maximum)
and (c) height, $\max|P|$, of the peak of $|P|$
on the inner shell at $t=30$ (Fig.~\ref{fig:sharp}(c)) against the grid spacing $dr$ at the shell, for
uniform grids (blue circles) and grids refined around the shell (red
squares), with a power-law fit for each group.}
\label{fig:wh}
\end{figure*}

On coarse grids the white-hole profiles
exhibit three narrow features, each with a different physical origin.
Figure~\ref{fig:sharp}(a) shows them on one profile, at $t=40$, and
Fig.~\ref{fig:wh}(a) follows their areal radius in time: the escaping
pulse (green), the inner shell (black) and the inner front (yellow). The part of the
scalar pulse that escapes is compressed as it moves outward through the
white-hole geometry, its width shrinking roughly as $e^{-0.15t}$. We
resolve it by adding a moving refinement to the coordinate map of
Sec.~\ref{sec:method}. The number of grid points does not change: the
computational coordinate is first reparametrised so that the points
cluster in a region, a few units wide, where the spacing is eight times
finer, and are drawn from a band around it, while the grid elsewhere is
left exactly as it was. This is not adaptive mesh refinement: there are
no nested grids, and the refinement factor is fixed. The centre of the
region follows the peak of $|P|$ through a feedback controller, and the
pulse is resolved as a thin but regular wave (Fig.~\ref{fig:sharp}(b)).
 Following it
further, however, would need a resolution growing exponentially in time.
The compression is visible in the outgoing light rays, which converge on
the pulse: the ray that leaves the inner shell at $t=30$, $3.2$ behind
the peak, is $0.3$ behind it at $t=34$ and $0.1$ at $t=40$
(Fig.~\ref{fig:wh}(a)).

The part of the pulse that does not escape
piles up, as an ingoing wave, on a sphere of areal radius
$\Theta\simeq6.9$, and at $t\simeq29.5$ it steepens beyond what any of our
grids resolves (Fig.~\ref{fig:sharp}(c)). Figure~\ref{fig:wh}(b,c) measures the width and
height of the peak of Fig.~\ref{fig:sharp}(c) at $t=30$ on seven
grids: three uniform, and four with a refined region of the same kind,
eight or sixteen times finer. Since the shell forms abruptly, with no
earlier peak to follow, this region is placed at the radius where the
uniform runs show the shell forming and is held there from $t=20$ until
$t=30.5$.
In each group of runs the width of the
peak shrinks roughly as $dr^{0.8}$, its height grows as $dr^{-0.5}$ and
the energy density as $dr^{-1}$.
This is consistent with a singularity forming at $t\simeq29.5$ but does
not prove it: the growth slows on the finest grids, and on the finest run
the width levels off once the refinement is centred on the shell, which may indicate a finite width of that order. A regular structure narrower than our finest
spacing cannot be excluded.

A third, weaker feature forms inside the shell. From $t\simeq7$ part of
the ingoing scalar pulse travels as a front, a jump of about $0.3$ in
$P$, along ingoing light rays, $\dot r=-\beta-N$. Where it travels the
lapse drops to $N\simeq0.13$, and the front nearly stalls at
$\Theta\simeq3.2$, close to the inner edge of the anti-trapped region
(Fig.~\ref{fig:wh}(a)). Its steepest slope changes little with resolution at
$t=24$ but doubles with each halving of $dr$ by $t=40$
(Fig.~\ref{fig:sharp}(d)), so it too
steepens beyond our resolution. The lapse then recovers, the front falls
to the centre at $t\simeq44$ and has dispersed by $t\simeq54$.

 No trapped surface exists
anywhere on the slice, and the outgoing light ray leaving the inner shell
joins the escaping pulse and travels outward with it
(Fig.~\ref{fig:wh}(a)).
We cannot exclude an event horizon at later times, but nothing in our data
suggests one. Beyond $t\simeq29.5$ the evolution does not converge: the
shell becomes a thin but resolved structure whose position depends on the
grid.
Every quantity behind the escaping pulse inherits this uncertainty.
In particular, the frozen centre seen in long runs on coarse
grids, where the central lapse collapses,
is
an artefact: the freezing time moves later on every finer grid
roughly as $dr^{-1/3}$, with no sign of converging.

 We repeated the amplitude exploration for
$(r_0,\sigma)=(10,3)$ and $(15,4)$, with the same outcome. Weak data
disperse ($A_0=0.01$, $0.015$ and $0.03$ for $\sigma=3$), intermediate data
form a transient white hole whose inner shell steepens as above,
and strong data form a black hole, with no steepening outside the
horizon. Near the black-hole threshold
the outcome depends on resolution, and coarse grids favour the white hole:
for $\sigma=3$, $A_0=0.08$ gives a trapped region that later disappears at
$dr\simeq0.009$, but a persistent black hole at $dr\simeq0.0047$. A trapped region
followed by an anti-trapped one does occur, for $r_0=15$, $A_0=0.028$:
the trapped region
is gone by $t\simeq20$ and the anti-trapped
region appears at $t\simeq24$, identically on three grids.
Finally, the collapse of the centre into a trapped region seen for
$\sigma=3$ is the frozen-centre artefact above: at $A_0=0.03$ it occurs at
$dr=0.0031$, not at $0.0016$, and again at $0.0008$, and at $A_0=0.04$ it
happens later on each finer grid
while its size shrinks with the spacing.

Our results broadly agree with earlier work. GEJ~\cite{Garfinkle:2007bk} already
reported, for a single $c_1=0.8$ run, that no horizon formed and ``the
evolution seems to become singular, thus indicating the formation of a
naked singularity''. Ref.~\cite{Akhoury:2016mrw} revisited the case with
higher resolution. We agree with the latter that at large $c_1$ a
moderately strong shell makes a temporary white hole and stronger shells
make black holes, but not on the fate.
Ref.~\cite{Akhoury:2016mrw} concludes that the fields disperse, noting that
the sharp features it finds near $t\simeq32$ are not fully resolved with
$40{,}000$ and $80{,}000$ points.
Those
spacings match our two coarsest uniform grids, on which the inner shell also
looks like a sharp but finite feature. Refining further shows
that it keeps steepening. Whether their feature is ours is not certain: their
shell, $(a_1/r)\,e^{-(r-r_0)^2/\sigma^2}$ with $\sigma=2$, is about $2.5$
times wider than ours, those runs use a different radial coordinate,
and their feature sits near the centre rather than at $\Theta\simeq7$. In the white-hole window, therefore, the evolution develops an unresolved
and possibly singular feature that no trapped surface hides, and beyond it
our runs do not determine the solution.

\section{Conclusion}
\label{sec:conclusion}

In spherical symmetry, Einstein--\ae ther theory and low-energy
Ho\v{r}ava gravity have the same collapse dynamics, in which both theories propagate a spin-0 mode. Spherical collapse is therefore
the cleanest nonlinear setting in which to ask whether the
spin-0 mode is a dynamical wave that carries energy away from the system. The code and the scripts behind every figure and number of this
paper are public~\cite{AEcollapse_code}.

Our simulations show that the spin-0 mode is indeed a wave. It is the
outgoing characteristic field of the evolution system, it carries the
whole part of the field that falls off as the inverse areal radius, and on
the dynamical black-hole background it propagates at the speed predicted
by the linear theory, also when that speed is superluminal. It carries a
finite, positive energy to infinity once the Coulomb-like \ae ther
acceleration field set up by the wave is subtracted; that field is linear
in the wave amplitude, dominates the Misner--Sharp luminosity and can be
identified with the surface term of the Einstein--\ae ther energy. For the superluminal coupling we studied, the collapse
radiates
$14\%$ of the ADM mass:
$9\%$ in the spin-0 wave and $5\%$ in the escaping scalar field.

These results do not fully settle the problem in more general settings, such as binary systems: Taherasghari and
Will~\cite{Taherasghari:2023rwn,Taherasghari:2025mlf} found a near-zone
dipole energy loss of compact binaries that depends only on the speed
$v_T$ of the transverse \ae ther waves, whereas the spin-0 speed also
enters the far-zone flux of
Refs.~\cite{Foster:2006az,Foster:2007gr,Yagi:2013ava}.
We have not computed the
dipole flux of a binary, and a spherical collapse, which has no
spin-1 mode and hence no $v_T$, cannot decide whether the near-zone
radiation-reaction force or the energy flux correctly describes
the energy loss of a two-body system. Within the spin-0 sector, however,
our results rule out that the mode is non-dynamical or carries no
energy. The energy leaves in a wave travelling at $v_0$, and
the Misner--Sharp mass lost through every sphere matches the \ae ther and scalar fluxes computed
independently from the stress tensor. If the spin-0 speed does drop out of the energy flux of a binary,
the reason must lie in the binary problem itself.

We also find that the static \ae ther left around the black hole has negative energy, so
the Misner--Sharp mass decreases outward from the horizon and the
dominant energy condition is violated. Inside the horizon the lapse
collapses and the fields freeze, as in the GEJ simulations
(Sec.~\ref{sec:method}). Ref.~\cite{Bhattacharyya:2015uxt} read this
freezing as the dynamical formation of a universal horizon, which does
form in the collapse of the version of Ho\v{r}ava gravity that survives
the GW170817 and solar-system bounds~\cite{Franchini:2021bpt}; we have not pursued this
identification. At large $c_1$ we recover the transient white holes of
Akhoury, Garfinkle and Gupta~\cite{Akhoury:2016mrw}, which the negative
energy of the \ae ther allows and GR does not. At higher resolution,
however, the inner shell that does not escape steepens beyond what any of
our grids resolves, with no trapped surface to hide it. This
may be the naked singularity suspected by GEJ~\cite{Garfinkle:2007bk} or
a regular structure finer than our grids. Our runs settle neither this
nor the solution beyond that point.

Three research directions follow. First, the energy measurement should be
repeated for couplings allowed by GW170817 and binary pulsars, for which
the spin-0 speed can differ substantially from the scalar speed, and
connected to the sensitivities of compact objects and to binary
observables~\cite{Foster:2006az,Ramos:2018oku,Barausse:2019yuk,Kidnew:2610.00503}. Second,
non-spherical simulations would reach the spin-1 sector, whose speed is
the one that appears in the radiation-reaction result of
Refs.~\cite{Taherasghari:2023rwn,Taherasghari:2025mlf}; the
characteristic decomposition of Sec.~\ref{sec:spin0} provides boundary
conditions and extraction variables for them. Third, the large-$c_1$
case calls for adaptive mesh refinement, whose resolution, unlike the
fixed factor used here, grows as the shell steepens, to decide whether a
naked singularity forms.
\begin{acknowledgments}
We acknowledge support from the European Union’s
Horizon ERC Synergy Grant “Making Sense of the
Unexpected in the Gravitational-Wave Sky” (Grant No.
GWSky–101167314). We used Claude (Anthropic) to help write, debug and run the numerical code, to prepare the figures and to edit the text; all results were checked by the authors, who take full responsibility for the content. We thank Ramiro Cayuso for helpful discussions and advice on the numerical methods used in this work.
\end{acknowledgments}

\appendix
\section{Convergence tests}
\label{sec:app-conv}

This appendix collects the self-convergence data behind
Sec.~\ref{sec:conv}, first for the scattering test and then for the
collapse test.

\begin{figure*}[t]
\centering
\maybegraphic[width=\textwidth]{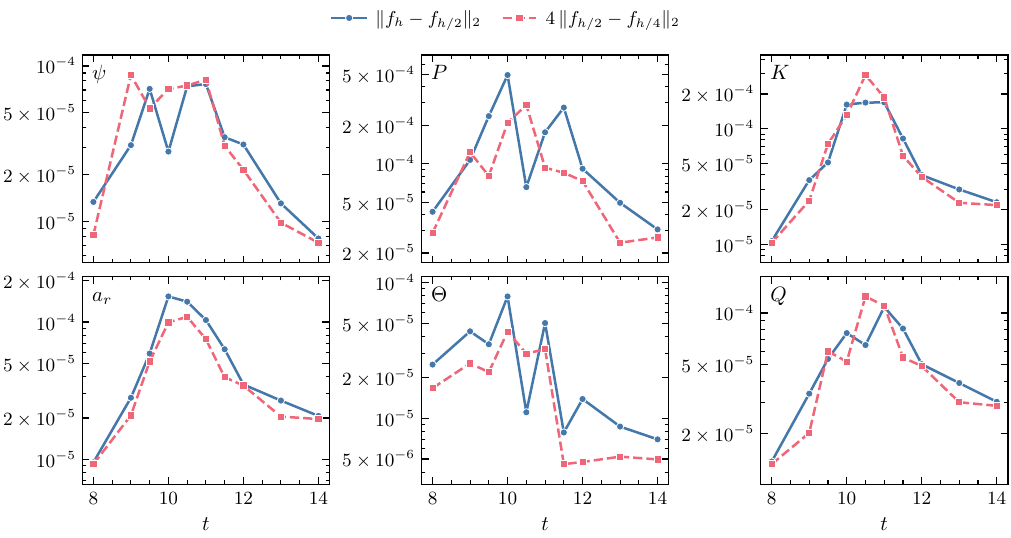}
\caption{Self-convergence of the evolved fields and $Q$ in scattering,
$A_0=0.01$, on the nested grids $N_r=6401/12801/25601$
($r_{\max}=150$).  Each
panel shows, against $t$ and on a logarithmic scale, the $L_2$ norms over
$0<r<20$ of $f_h-f_{h/2}$ 
and $4(f_{h/2}-f_{h/4})$. The curves overlap for
second-order convergence.}
\label{fig:conv-fields-A001}

\vspace{\baselineskip}

\maybegraphic[width=\textwidth]{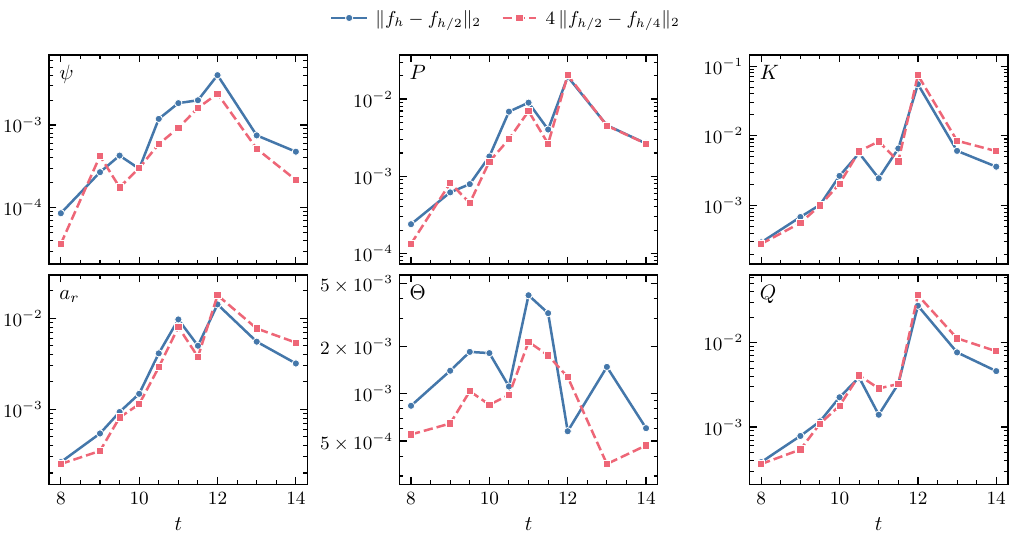}
\caption{As Fig.~\ref{fig:conv-fields-A001}, for $A_0=0.06$, the
strongest shell of the scattering test. The curves of $\trK$, $a_r$ and
$Q$ separate after the bounce ($t\gtrsim12$).}
\label{fig:conv-fields-A006}
\end{figure*}

\begin{figure*}[!t]
\centering
\maybegraphic[width=\textwidth]{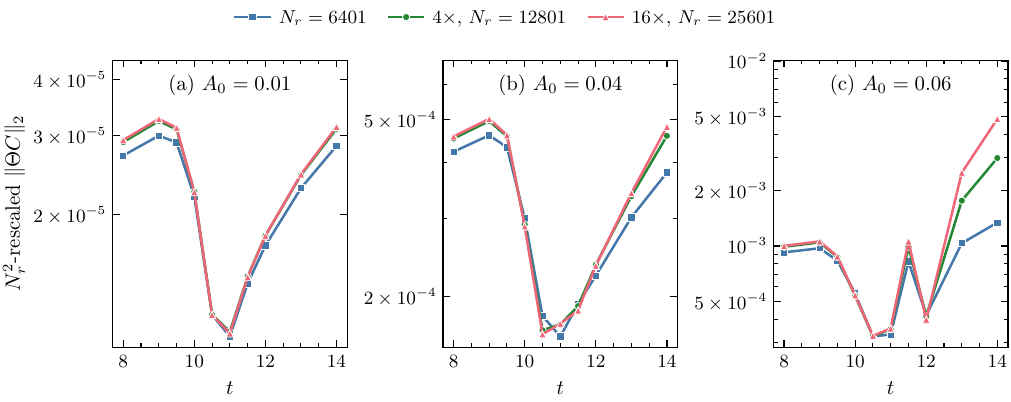}
\caption{Hamiltonian constraint in scattering: $\|\Theta C\|_2$ over $0<r<20$ against
$t$ for $N_r=6401/12801/25601$, rescaled by $N_r^2$ and normalised to
the coarsest grid (factors $1$, $4$ and $16$). (a) Weak field, $A_0=0.01$; (b) strong field, $A_0=0.04$;
(c) $A_0=0.06$, the strongest shell. In the weak and
strong-field cases the three resolutions collapse onto one curve,
i.e.\ $\|\Theta C\|\to0$ at second order; for $A_0=0.06$ they
separate after the bounce ($t\gtrsim12$).}
\label{fig:conv-constraint}
\end{figure*}

\emph{Scattering.} Figures~\ref{fig:conv-fields-A001}
and~\ref{fig:conv-fields-A006} show the differences of the evolved fields
and of $Q$ between successive grids for the weakest and the strongest
shell. For $A_0=0.01$ the two curves overlap, as second-order convergence
requires, except at isolated times during the bounce, when a difference
passes through an accidental minimum; for $A_0=0.06$ the curves of the
\ae ther variables separate after the bounce. Figure~\ref{fig:conv-constraint}
shows the Hamiltonian constraint for the three amplitudes of
Table~\ref{tab:conv-scatter}.

\begin{figure}[!htb]
\centering
\maybegraphic[width=\columnwidth]{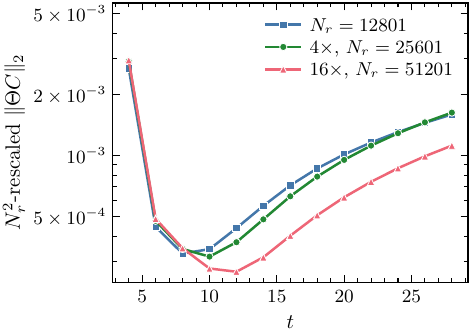}
\caption{Hamiltonian constraint in the collapse test ($A_0=0.35$,
$r_{\max}=300$, fixed grid): $\|\Theta C\|_2$ outside the apparent
horizon, over $r_{\rm AH}+2\le r\le150$, against $t$ on the nested grids
$N_r=12801/25601/51201$, rescaled by $N_r^2$. The rescaled curves
coincide if $\|\Theta C\|\to0$ at second order. From $t\simeq10$ the
curve of the finest grid lies below the other two: on the finer pair of
grids the constraint converges faster than second order.}
\label{fig:conv-collapse}
\end{figure}

\emph{Collapse.} The constraint-norm ratios quoted in
Sec.~\ref{sec:conv} are listed in Table~\ref{tab:conv-ratio}, and the
rescaled constraint norm is shown in Fig.~\ref{fig:conv-collapse}. The
ratios must be read on the nested \emph{fixed} grid: with the moving
mesh the three resolutions are no longer coincident point by point, and
the stretched outer zone degrades the comparison. The apparent horizon
is present from $t=2$, with $\ThAH\simeq8.5$, and by $t=28$ the central
lapse is below $10^{-6}$. Because the grid falls inward, the window of
the norms drifts: $r_{\rm AH}$ grows from $\simeq9$ at $t=2$ to $\simeq65$
at $t=28$, while the areal radius at $r=150$ decreases from $\simeq137$
to $\simeq90$. At $t=28$, the areal radius of the apparent horizon is
$\ThAH=12.293$, $12.604$ and $12.677$ on the three grids. On the moving
mesh at $N_r=51201$, the ADM mass read at $\Theta=250$ is $3.70836$ at
$t=2$ and $3.71107$ at $t=28$.
\begin{table}[h]
\centering
\caption{Constraint-norm ratios outside the apparent horizon
($r_{\rm AH}+2\le r\le150$) on the nested fixed grid
$N_r=12801/25601/51201$ ($r_{\max}=300$, $A_0=0.35$); a ratio of $4$ is
exact second order.}
\label{tab:conv-ratio}
\begin{tabular}{lccccccc}
\toprule
$t$ & 4 & 8 & 12 & 16 & 20 & 24 & 28\\
\midrule
$\|\Theta C\|_{12801}/\|\Theta C\|_{25601}$ & 3.69 & 3.80 & 4.69 & 4.53 & 4.26 & 4.05 & 3.92\\
$\|\Theta C\|_{25601}/\|\Theta C\|_{51201}$ & 3.95 & 3.97 & 5.59 & 6.26 & 6.12 & 5.95 & 5.82\\
\bottomrule
\end{tabular}
\end{table}

\section{Non-local field redefinitions and the spin-0 mode}
\label{app:pn_redefinition}
Below we first give a simple example of the mechanism described in
Sec.~\ref{sec:pn_comparison}, using a single propagating scalar, and
then derive the results quoted there for the spherically symmetric
scalar sector of the PN-expanded equations of
Ref.~\cite{Taherasghari:2023rwn}. Throughout, $\eta$ counts PN orders,
a tilde denotes the original variables, and untilded symbols denote the
variables obtained after the field redefinition.

\subsection{A single-field example}
\label{app:pn_toy}

Consider a scalar field $\tilde\phi$ propagating with speed $v$,
\begin{equation}
\left(\nabla^2-\frac{\eta}{v^2}\partial_t^2\right)\tilde\phi=0,
\label{eq:simple_scalar}
\end{equation}
where $\partial_t=O(1)$ in the adopted
counting. On a mode $e^{-i\omega t}f(r)$ with $\nabla^2 f=-k^2f$, the
operator acts as multiplication by
\begin{equation}
\mathcal{L}=-k^2(1-y),\quad
y\equiv\frac{\eta\omega^2}{v^2k^2}
\;\longleftrightarrow\;
\frac{\eta}{v^2}\partial_t^2\nabla^{-2}.
\label{eq:simple_L}
\end{equation}
The dependence on $y$ encodes the time-dependent propagation
correction: each additional power of $y$ carries one more PN order and
one more inverse Laplacian.

Consider now the non-local field redefinition $\tilde\phi=T_p\phi$, which
expresses the original field $\tilde\phi$ in terms of the redefined
field $\phi$, with
\begin{equation}
T_p=\sum_{j=0}^{p}y^j.
\label{eq:Tp}
\end{equation}
The transformed operator is
\begin{equation}
\mathcal{L}'=\mathcal{L}T_p=-k^2(1-y)\sum_{j=0}^{p}y^j
=-k^2(1-y^{p+1}).
\label{eq:Lprime}
\end{equation}
The propagation correction is therefore absent through $p$PN and first
appears at $(p+1)$PN, through a term proportional to
$\eta^{p+1}\partial_t^{2p+2}(\nabla^{-2})^p$. For $p=1$, for
example,
\begin{equation}
\mathcal{L}'=-k^2+\frac{\eta^2\omega^4}{v^4k^2}.
\label{eq:Lprime_p1}
\end{equation}
The physical root $y=1$ is still a root of $\mathcal{L}'$. The polynomial
transformation, however, also introduces the roots $y^{p+1}=1$ with
$y\ne1$; these are artefacts of the transformation and must not be
mistaken for additional physical modes. As for any field redefinition,
statements about the mode content hold only where the transformation is
well defined and invertible.

Thus, a finite PN truncation of a non-local field redefinition can
postpone the explicit appearance of propagation effects without
implying that the underlying degree of freedom has disappeared. The
rest of this appendix shows that the same mechanism operates in the
coupled scalar sector of Einstein--\ae ther theory.

\subsection{Spherical reduction and the linear scalar mode}
\label{app:pn_reduction}

We reduce Eqs.~(4.5)--(4.6) of Ref.~\cite{Taherasghari:2023rwn} to spherical symmetry,
using
\begin{equation}
\begin{gathered}
\tilde K^r=\partial_r\tilde\kappa,\qquad
\tilde K^r_{\text{\ae}}=\partial_r\tilde\kappa_{\text{\ae}},\\
\tilde B^{jk}=\delta^{jk}\tilde\gamma_1+\partial_j\partial_k\tilde\gamma_2,
\end{gathered}
\end{equation}
so that $\tilde B^{rr}_{\rm LTF}=\tfrac23(\tilde\gamma_2''-\tilde\gamma_2'/r)$. Since
\begin{equation}
\left(\nabla^2-\frac{2}{r^2}\right)\partial_r\tilde\kappa
=\partial_r\nabla^2\tilde\kappa,
\end{equation}
the $1/r^2$ terms in the vector and tensor Laplacians are absorbed, and
every operator in Eqs.~(4.5)--(4.6) becomes a function of $\nabla^2$
and $\partial_t$ acting on the scalar potentials. For example,
Eqs.~(4.5a) and (4.6) become
\begin{align}
&(1-\tfrac12c_{14})\nabla^2\tilde N-\eta\ddot{\tilde N}
-\tfrac12c_{14}\eta\nabla^2\tilde B
\nonumber\\
&\qquad
+2\eta c_{14}\nabla^2\partial_t(\tilde\kappa_{\text{\ae}}-\tilde\kappa)=0,
\\
&-\tfrac14c_{14}\big[\dot{\tilde N}+\eta\dot{\tilde B}
-4\eta(\ddot{\tilde\kappa}_{\text{\ae}}-\ddot{\tilde\kappa})\big]
\nonumber\\
&\qquad
-\tfrac14c_2\big[3\dot{\tilde N}+4\nabla^2\tilde\kappa_{\text{\ae}}
-\eta\dot{\tilde B}\big]=0,
\end{align}
and the harmonic-gauge conditions read
\begin{equation}
\dot{\tilde N}+\nabla^2\tilde\kappa=0,\qquad
\dot{\tilde\kappa}+\tilde\gamma_1+\nabla^2\tilde\gamma_2=0.
\end{equation}
The remaining reduced equations are obtained in the same way.

On a mode satisfying $\nabla^2 f=-k^2f$ (for example,
$f=\sin(kr)/r$), one has $\nabla^{-2}\to-1/k^2$. In analogy with
Eq.~\eqref{eq:simple_L}, we introduce
\begin{equation}
x\equiv\frac{\eta\omega^2}{k^2}
\quad\longleftrightarrow\quad
\eta\partial_t^2\nabla^{-2},
\label{eq:x_def}
\end{equation}
so that $\Box\to-k^2(1-x)$; each power of $x$ represents one PN order,
together with one inverse Laplacian. With the rescaled amplitudes
\[
\tilde K=(\omega/k)\hat{\tilde K},\qquad
\tilde K_{\text{\ae}}=(\omega/k)\hat{\tilde K}_{\text{\ae}},\qquad
\tilde B=(\omega/k)^2\hat{\tilde B},
\]
the gauge relation gives $\hat{\tilde K}=\tilde N$; in what follows we
drop the hats on $\hat{\tilde K}_{\text{\ae}}$ and $\hat{\tilde B}$. The
three independent equations, Eqs.~(4.5a), (4.5c)$_{\rm tr}$, and (4.6)
of Ref.~\cite{Taherasghari:2023rwn}, for $(\tilde N,\tilde B,\tilde K_{\text{\ae}})$ then
take the form
\begin{equation}
\mathcal{D}_0(x)(\tilde N,\tilde B,\tilde K_{\text{\ae}})^T=0,
\end{equation}
with
\begin{align}
&\mathcal{D}_0=\\
&{\setlength{\arraycolsep}{3pt}\medmuskip=3mu
\begin{pmatrix}
-1+\tfrac12c_{14}-(2c_{14}-1)x
& \tfrac12c_{14}x & 2c_{14}x\\[2pt]
-\tfrac92c_2
& -1+(1+\tfrac32c_2)x & 6c_2\\[2pt]
-\tfrac14(3c_2+c_{14})+c_{14}x
& \tfrac14(c_2-c_{14})x & c_2-c_{14}x
\end{pmatrix}}.\nonumber
\end{align}
Its determinant is
\begin{equation}
\det \mathcal{D}_0=\tfrac12(2-c_{14})c_2(1-x)^2
\left(1-\frac{x}{v_L^2}\right),
\label{eq:det_M0}
\end{equation}
where, for $c_{13}=0$,
\begin{equation}
v_L^2=\frac{c_2(2-c_{14})}{c_{14}(2+3c_2)}.
\label{eq:vL_def}
\end{equation}
The factors $(1-x)$ correspond to gauge/constraint modes, while the
physical scalar mode is associated with the root $x=v_L^2$.

Eliminating $\tilde B$ and $\tilde K_{\text{\ae}}$ yields the reduced equation
$s_{-1}(x)\tilde N=0$,
with
\begin{equation}
\begin{aligned}
s_{-1}(x)
&=-\left(1-\tfrac12c_{14}\right)
\left[1-(1+g_1)x+O(x^2)\right],
\end{aligned}
\label{eq:sminus1}
\end{equation}
where
\begin{equation}
g_1\equiv\frac{c_{14}}{2v_L^2}.
\end{equation}
The term linear in $x$ splits into the $\Box\tilde N$ contribution of the
diagonal equation (4.13a) of Ref.~\cite{Taherasghari:2023rwn} and an additional
$v_L$-dependent 1PN term,
\begin{equation}
\left(1-\tfrac12c_{14}\right)g_1x
=\frac{c_{14}(2-c_{14})}{4v_L^2}x.
\label{eq:pn_1_term}
\end{equation}

\subsection{First-order field redefinition and its interpretation}
\label{app:pn_first_order}

We now apply Eq.~(4.7) of Ref.~\cite{Taherasghari:2023rwn}, truncated at first order in
$\eta$. For the lapse, the transformation expresses the original
variable in terms of the redefined ones as
\begin{equation}
\begin{aligned}
\tilde N
&=N+\frac{\eta c_{14}}{2-c_{14}}B
+\frac{\eta}{v_L^2}\dot R
\\
&=\left[1+\frac{c_{14}}{2v_L^2}
\eta\partial_t^2\nabla^{-2}\right]N
\\
&\quad
+x\left[\frac{c_{14}}{2-c_{14}}B
+\frac{2c_1}{v_L^2}K_{\text{\ae}}\right],
\end{aligned}
\label{eq:N_transform}
\end{equation}
together with the corresponding first-order expressions for
$\tilde K$, $\tilde B$, and $\tilde K_{\text{\ae}}$. Writing
\[
(\tilde N,\tilde B,\tilde K_{\text{\ae}})^T=T_1(x)(N,B,K_{\text{\ae}})^T,
\]
the original equations $\mathcal{D}_0(\tilde N,\tilde B,\tilde K_{\text{\ae}})^T=0$
become $\mathcal{D}_1(N,B,K_{\text{\ae}})^T=0$ for the redefined variables, with $\mathcal{D}_1=\mathcal{D}_0T_1$. At
order $x^0$, the first
column of $T_1$ is
\begin{equation}
\left(1,\tfrac32c_{14},
\frac{c_{14}+3c_2}{4c_2}\right)^T,
\end{equation}
which cancels the zeroth-order response of $(B,K_{\text{\ae}})$ to $N$. Hence
\begin{equation}
\begin{split}
\mathcal{D}_1(x)={}&\mathrm{diag}\Bigl(-(1-\tfrac12c_{14})(1-x),\,-(1-x),\\
&\qquad c_1(1-x/v_T^2)\Bigr)+O(x^2),
\end{split}
\label{eq:M1}
\end{equation}
where $v_T^2=c_1/c_{14}$ is the spin-1 speed for $c_{13}=0$. These are the symbols of the diagonal
operators $(1-\tfrac12c_{14})\Box N$, $\Box B$, and
$-c_1\Box^*K_{\text{\ae}}$ in Eqs.~(4.13a,d,e) of Ref.~\cite{Taherasghari:2023rwn}. The
coefficient of $x$ in the lapse equation is now $(1-\tfrac12c_{14})$,
independent of $c_2$ and hence of $v_L$: the term in
Eq.~\eqref{eq:pn_1_term} has been removed.

At the next order, however, the reduced operator acting on the
redefined lapse $N$ is
\begin{equation}
s_1(x)=-(1-\tfrac12c_{14})(1-x)+\kappa_1x^2+O(x^3),
\label{eq:s1}
\end{equation}
with
\begin{equation}
\begin{aligned}
\kappa_1
&=g_1\frac{c_{14}(2+3c_2)}{2c_2}
\\
&=\frac{c_{14}(2-c_{14})}{4v_L^4}.
\end{aligned}
\label{eq:kappa1}
\end{equation}
The first $v_L$-dependent term in the transformed lapse equation is
therefore proportional to
\begin{equation}
\kappa_1\eta^2\partial_t^4\nabla^{-2}N,
\end{equation}
which is 2PN in the adopted counting.

The origin of this term follows from the exact relation
\eqref{eq:G_exact} between $s_{-1}$, which acts on the original lapse
$\tilde N$, and the diagonal lapse operator, with $y=x/v_L^2$. Expanding
the inverse propagator \eqref{eq:inverse_propagator} in PN orders gives
\begin{equation}
\Gamma=1+\frac{c_{14}}2(y+y^2+\cdots).
\end{equation}
The first-order redefinition retains only the first correction in this
series, so the first neglected power contributes at 2PN. In the
notation above,
\begin{equation}
s_1=s_{-1}\left(1+\frac{c_{14}}2y\right)+O(x^3).
\end{equation}
This is the multi-field counterpart of Eq.~\eqref{eq:Lprime}, whose
operator form is
\begin{equation}
\begin{split}
&\left(\nabla^2-\frac{\eta}{v^2}\partial_t^2\right)
\sum_{j=0}^{p}
\left(\frac{\eta}{v^2}\partial_t^2\nabla^{-2}\right)^j
\\
&\qquad=\nabla^2\left[1-
\left(\frac{\eta}{v^2}\partial_t^2\nabla^{-2}\right)^{p+1}
\right],
\end{split}
\end{equation}
understood on modes for which the inverse Laplacian is defined. A
finite PN truncation can therefore shift the order at which the
propagation-speed dependence appears explicitly in a chosen equation,
whereas a local redefinition, without inverse Laplacians, cannot
generate the same series of powers of $\partial_t^2\nabla^{-2}$.

As in the single-field case, the mode content is preserved only where
the transformation is invertible. For a finite transformation matrix
$T_1$, the identity
\begin{equation}
\det \mathcal{D}_1=\det \mathcal{D}_0\,\det T_1
\end{equation}
shows that the physical root of $\det \mathcal{D}_0$ is retained whenever
$\det T_1$ is nonzero there, while additional roots may appear where
$\det T_1=0$. The exact factor $\Gamma(x)$, on the other hand, is singular
at the physical pole $x=v_L^2$ and does not correspond to an invertible
redefinition there (see Sec.~\ref{sec:pn_comparison}).

In summary, in the first-order truncation the explicit $v_L$
dependence of the reduced lapse equation is absent through 1PN and
reappears at 2PN. Retaining higher orders in the redefinition is
expected to push it to higher PN order, as discussed in
Sec.~\ref{sec:pn_comparison}.

\end{document}